\RequirePackage{xcolor}
\documentclass[hyper]{JHEP}
\usepackage{graphicx,amssymb,bm,latexsym,amsmath}
\usepackage{epstopdf}
\usepackage{caption}
\usepackage{subcaption}
\newcommand{\bej}[1]{ \begin{equation}\label{#1} }
\newcommand{\eej}{\end{equation}}
\newcommand{\beaj}[1]{\begin{eqnarray}\label{#1} }
\newcommand{\eeaj}{\end{eqnarray}}
\newcommand{\eq}[1]{(\ref{#1})}
\def\ZZZ{{\hskip-3pt\hbox{ Z\kern-1.6mm Z}}}
\def\zzz{{\hskip-3pt\hbox{ z\kern-1mm z}}}

\newcommand{\bd}{\bar{\rm D}}

\newcommand{\N}{\frac{m_{2}}{k_{2}}-\frac{m_{1}}{k_{1}}}

\newcommand{\be}{\begin{equation}}
\newcommand{\ee}{\end{equation}}
\newcommand{\ben}{\begin{eqnarray}\displaystyle}
\newcommand{\een}{\end{eqnarray}}

\def\one{{\hbox{ 1\kern-.8mm l}}}
\def\zero{{\hbox{ 0\kern-1.5mm 0}}}

\def\be{\begin{equation}}       
\def\ee{\end{equation}}         

\def\bea{\begin{eqnarray}}      
\def\eea{\end{eqnarray}}

\def\ba{\begin{array}}
	\def\ea{\end{array}}
\def\bd{\begin{displaymath}}
\def\ed{\end{displaymath}}

\def\eq{\begin{equation}}
\def\eqe{\end{equation}}
\def\eqa{\begin{eqnarray}}
\def\eqae{\end{eqnarray}}
\def\ena{\end{eqnarray}}

\def\unit{1 \hskip-.3em \raise2pt\hbox{$ \scriptstyle |$ } }
\def\bd{\begin{displaymath}}
\def\ed{\end{displaymath}}

\def\6{\partial}
\def\N4{{\cal N}=4}

\def\bop#1{\setbox0=\hbox{$#1M$}\mkern1.5mu
\vbox{\hrule height0pt depth.04\ht0
	\hbox{\vrule width.04\ht0 height.9\ht0 \kern.9\ht0
		\vrule width.04\ht0}\hrule height.04\ht0}\mkern1.5mu}
\def\>{\rangle} %right angle

\def\<{\langle} %left angle
\def\Dsl{D \hskip-.6em \raise1pt\hbox{$ / $ } }
\def\to{\rightarrow}

\def\+{\oplus}

\def\as2{AdS_3\times S^3_1 \times S^3_2}

\title{On N-spike strings in conformal gauge with NS-NS fluxes }
\author{Aritra Banerjee$^1$, Sagar Biswas$^2$, Priyadarshini Pandit$^3$, Kamal L. Panigrahi$^3$ \\
$^1$CAS Key Laboratory of Theoretical Physics,\\ Institute of Theoretical Physics, Chinese Academy of Sciences,\\  Beijing 100190, China\\
$^2$Department of Physics, Ramakrishna Mission Vidyamandira, \\Belur Math, Howrah 711 202, India\\
$^3$ Department of Physics, Indian Institute of Technology Kharagpur,\\ Kharagpur 721 302, India\\

Email: \email{aritra@itp.ac.cn,biswas.sagar@vidyamandira.ac.in,pandit006@iitkgp.ac.in panigrahi@phy.iitkgp.ac.in}}

\abstract{The $AdS_3\times S^3$ string sigma model supported both by NS-NS and R-R fluxes has become a well known integrable model, however a putative dual field theory description remains incomplete. We study the anomalous dimensions of twist operators in this theory via semiclassical string methods. We describe the construction of a multi-cusp closed string in conformal gauge moving in  $AdS_3$ with fluxes, which supposedly is dual to a general higher twist operator. After analyzing the string profiles and conserved charges for the string, we find the exact dispersion relation between the charges in the `long' string limit. This dispersion relation in leading order turns out to be similar to the case of pure RR flux, with the coupling being scaled by a factor that depends on the amount of NS-NS flux turned on. We also analyse the case of pure NS flux, where the dispersion relation simplifies considerably. Furthermore, we discuss the implications of these results at length. }

\keywords{Integrability, AdS/CFT correspondence}

\begin{document}
\section{Introduction}
	The AdS/CFT correspondence \cite{a} , which recently completed two decades of glorious presence in all of theoretical physics, in its purest form states that string theory in bulk $d$ dimensional $AdS$ background is equivalent to a conformal field theory living on the $(d - 1)$-dimensional boundary space.
This revelation immediately opened up a fascinating possibility to solve the non trivial strongly coupled  dual field theory, at least in the planar limit, where the duality works exactly. The crux of the duality in most elementary terms relates various observables in the boundary field theory, such as anomalous dimensions of trace operators to the energy spectrum of dual string states. Consequently, the spectral problem of AdS/CFT has become one of the most sought after problems in high energy physics. This was further simplified due to presence of \textit{integrability} on both sides of the duality, such as in the well known case of $AdS_5/CFT_4$.
 The remarkable observation that the dilatation operator of the  $4d$ Super Yang-Mills(SYM) theory at
one loop can be written exactly as the Hamiltonian of an integrable Heisenberg spin-chain has
been a key concept in connecting integrability, strongly coupled systems and string
theory. On the other hand, dual string dynamics in $AdS_5\times S^5$ is described by an integrable non-linear sigma model, providing us with exact grasp on the spectrum of the theory. This also means, in the classical limit, anomalous dimensions of certain operators in the dual SYM theory can be extracted from the dispersion relation between conserved ``large'' charges of certain integrable string motion in $AdS$.

One of the most interesting cases of such a classical string picture might be that
of the folded spinning string or GKP string found in \cite{GKP}. This case relates anomalous dimensions of certain operators in the $SL(2)$ sector of the SYM theory to energy-spin relation of closed strings spinning in $AdS_3\subset AdS_5$. 
Operators in the $SL(2)$ sector are constructed from a single complex scalar field $\Phi$, and a covariant derivative $\mathcal{D}$. A general single trace operator then takes the form,
$$ \mathcal{O} = \text{Tr}( \mathcal{D}^S \Phi^L). $$
Here $L$ is the length or twist of the operator and $S$ is the Lorentz spin. These operators are of particular interest in both string theory and quantum field theory, especially important in a twist-two operator $ \mathcal{O}_{\Delta} = \text{Tr}(\Phi \mathcal{D}^S \Phi)$, where perturbative computations in the field theory gives the anomalous dimension as,
$$ \Delta-(S+2)=f(\lambda)\log~S+\mathcal{O}\left( \frac{1}{S}\right), $$ where $\lambda$ is the well known 't Hooft coupling constant. 
The interpolating function in weak coupling $f(\lambda)\sim \lambda$ appears in the expectation value of a straight Wilson loop with a ``cusp'' and is
therefore often referred to as the \textit{cusp anomalous dimension}. The exact dual object is the closed folded strings spinning in $AdS_3\subset AdS_5$, whose segments pass through the center $\rho=0$ in global $AdS$ coordinates. The energy of such a string can be calculated from the associated strong coupling sigma model solution in the large spin ($S \to \infty$) limit\footnote{The large spin or classical limit of the solution, where the logarithmic scaling appears in anomalous dimension, is often denoted as $S\to \infty$, $\sqrt{\lambda}\to \infty$, together with $\frac{S}{\sqrt{\lambda}}\gg 1$.}, and reads,
$$  	E-S=\frac{\sqrt{\lambda}}{\pi}\log S,~~~~~\sqrt{\lambda}=\frac{R^2}{\alpha^\prime}\gg1,$$
	here $R$ is the radius of $AdS$ and $\alpha^\prime=\frac{1}{2\pi T}$ is the string tension parameter. This string however is a special case of a spinning spiky string, which is related to twist operators with derivatives on several fields, where the derivative contribution dominates in the large spin limit.
 It is shown in \cite{spiky1} that a general higher twist operator Tr$(\mathcal{D}^{S_1}\Phi_1 \mathcal{D}^{S_2}\Phi_2...\mathcal{D}^{S_N}\Phi_N)$ is dual to the string solution with $N$ cusps or spikes. As expected, when $N$ is taken to be 2, the dispersion relation exactly reduces to the folded spinning string  or GKP string. From the worldsheet perspective, these spikes (or
cusps) are defined as a derivative discontinuity in the unit space-like tangent vector on the $2d$ worldsheet,
which although does not spoil the criterion of a smooth worldsheet. A number of studies in such $N$-cusped strings in $AdS$ and related spaces have followed , including e.g. \cite{spiky1}-\cite{Banerjee:2015nha}.

Although $AdS_5/CFT_4$ is the most successful example of the duality, a number of lower dimensional versions have also been very intensively studied in the available literature. Superstring theory on $AdS_3 \times S^3\times \mathcal{M}^4$ backgrounds have recently attracted renewed interest in the context of studies in $AdS_3/CFT_2$ correspondence \cite{b}\footnote{For a recent review of this vast subject, we redirect the reader to to excellent introduction \cite{Sfondrini} and references therein.}. The most tractable cases of this duality with the compact manifold $\mathcal{M}^4$ being $T^4$ \cite{VIII}-\cite{BPS} or $S^3\times S^1$ \cite{SS}-\cite{EberhardtGopakumar} have played a crucial role in understanding the duality. Added advantage is that for both cases the string sigma models have been proven to be classically integrable \cite{BSZ,wulff} \footnote{See \cite{review} for a general introduction to integrability in string sigma models.}. Not only the case of $AdS_3 \times S^3\times T^4$ with RR fluxes, the sigma model for the same supported by both NS-NS and R-R three-form fluxes has also been proven to be classically integrable  \cite{CagnazzoZarembo}. The background solution can be shown to satisfy 
IIB supergravity field equations, provided the parameters associated to field strength of NS-NS fluxes $q$ and the one associated to the strength of R-R fluxes (say $\hat q$) are related by the constraint $q^2+\hat q^2=1$. Depending on the value of parameters  the string sigma model interpolates between that of the SL(2, R) WZW model \cite{Moo} (pure NS-NS) and the one supported only by RR fluxes. 

A lot of discussion has appeared on this particular model supported by both form of fluxes, including fundamentals of integrable nature and S-matrix \cite{BianchiHoare},\cite{HT}-\cite{Modulispace}. Moreover  a large number of classical string solutions in this string background has appeared in last few years, mainly focussing on how turning on the flux changes the solutions and hence the nature of possible dual observables. All of it started as the  `Giant Magnon' dispersion relation was proposed in \cite{B.Hoare}, which turned out to have non-periodic contribution in the momentum as the periodicity is explicitly spoiled by the flux. Subsequently finite-gap solutions were proposed in \cite{FinitegapBabichenko,masslessfinitegap} and a large class of folded \cite{Rotating1}, rotating \cite{both, C.Ahn, Banerjee:2018goh}, pulsating \cite{Rotating2, Barik1} and GKP-like multi-spike \cite{multispike, stepanchuk} strings have been studied in the literature. Also, well known Neumann-Rosochatius dynamical model of strings has been generalised to $AdS_3\times S^3$ with NS-NS fluxes together with relevant classical solutions \cite{HN1,HN2,HN3} and one-loop quantization \cite{HN4} being elucidated in a large number of papers. Moreover, probe D1 strings exploring the mixed flux background have been discussed in \cite{D1}.  Minimal surfaces for this background has been discussed in detail too \cite{Rotating1,Hernandez:2018lvh}.

	One of the main explorations in this connection has been about the analogue of the folded string in this background, and subsequently the twist-operators in a putative dual theory. The first discussion of this appeared in \cite{Rotating1}, where the solution was constructed by solving the dynamical equation and then finding the dispersion relation between the charges using a perturbative expansion of in the flux parameter $q$. Curiously, the solution had some subtleties in imposing closed string boundary condition and it was found that the next to leading order cusp anomalous dimension (upto $\mathcal{O}(q^2)$) in that case starts with a term containing $-\frac{q^2}{2}(\log~S)^2$. In the paper \cite{multispike} the problem was revisited using a N-spike Kruczenski (static gauge) string solution and the dispersion was again calculated perturbatively in $q$, where the next to leading order correction turns out  $-\frac{q^2}{2}\log~S$, at complete odds with  \cite{Rotating1}. Finally in \cite{stepanchuk} it was conjectured using a worldsheet transformation between $q=0$ and $q\neq 0$ strings in $AdS_3\times S^1$, that the dispersion relation in the latter case remains the same as in the pure RR case, but the logarithmic divergence is scaled by a factor of $\sqrt{1-q^2}$.

	In the current article we want to give a definitive answer about the dispersion relation via an exact calculation in $q$. We have again concentrated in this paper on the N-spike string solution in the mixed flux background, but this time following the same procedure used in \cite{Jevicki:2008mm, JJ}, i.e. via a computation using the sigma model in conformal gauge. We find exact agreement with the observations of  \cite{stepanchuk}, as we explicitly get the dispersion relation to be,
	\begin{equation}
	E-S=N\frac{\sqrt{\lambda}}{\pi}\sqrt{1-q^2}~\left[\frac{1}{2}\log S+...\right].
	\end{equation}
	Here putting number of spikes $N=2$ gets one back to the desired relation for the folded string in this background. We also discuss the structure and string profile associated to the solution in this work. Furthermore, we discuss the fate of these solutions and the scaling relation among various conserved charges in case of  pure NS-NS flux i.e. for $q=1$. This limit considerably simplifies the solutions and appears to be quite subtle to take. 
	
	The organisation of this paper is as follows. Section 2 deals with the revisiting of the N-soliton solution in conformal gauge which includes the string solution, string boundary condition and dispersion relation along charges. In section 3, we repeat the corresponding computations and find N-Soliton solution in presence of mixed flux. Associated modified boundary conditions and the changed dispersion relation has been elucidated here along with
	the case of pure NS-NS flux ($q=1)$.  Finally, in section 4 we present our conclusions and outlook. 
		
	\section{Revisiting the N-soliton solution in conformal gauge}
	
	In this section we provide a short review of the construction of multi-spike string in conformal gauge. This would serve as our base as we move along and do the similar calculation in presence of flux in the next section. We would be closely following the conventions and notations of \cite{Jevicki:2008mm} throughout.
	\subsection{The solution}
	In a general background, we can study a fundamental string coupled to an antisymmetric NS-NS B-field.  We start with the Polyakov action with a Wess-Zumino term
	
	\begin{equation}
	S=-\frac{\sqrt{\lambda}}{4\pi}\int{d\sigma d\tau\left[\sqrt{-\gamma}\gamma^{\alpha\beta} g_{MN}\partial_\alpha{X^M}\partial_\beta{X^N}-\epsilon^{\alpha\beta}\partial_\alpha{X^M}\partial_\beta{X^N}b_{MN}\right]}.
	\end{equation} 
	where $\lambda$ is t'Hooft coupling, $\gamma^{\alpha\beta}$ is the worldsheet metric and $\epsilon^{\alpha\beta}$ is the anti symmetric tensor defined as  $\epsilon^{\tau\sigma} =- \epsilon^{\sigma\tau}=1$. Variation of the action with respect to $X^M$ give us the following equations of motion,
	\begin{equation}
	\begin{split}
	2\partial_\alpha(\eta^{\alpha\beta}\partial_\beta{X^N}g_{KN})-&\eta^{\alpha\beta}\partial_\alpha{X^M}\partial_\beta{X^N}\partial_Kg_{MN}-2\partial_\alpha(\epsilon^{\alpha\beta}\partial_\beta{X^N}b_{KN})\\&+
	\epsilon^{\alpha\beta}\partial_\alpha{X^M}\partial_\beta{X^N}\partial_K b_{MN}=0
	\end{split}
	\end{equation}
	and variation with respect to the metric gives us the two Virasoro constraints,
	\begin{equation}
	 g_{MN}(\partial_\tau{X^M}\partial_\tau{X^N}+\partial_\sigma{X^M}\partial_\sigma{X^N})=0;
	~~~
	   g_{MN}(\partial_\tau{X^M}\partial_\sigma{X^N})=0.
	\end{equation}
	We use the conformal gauge (i.e.$\sqrt{-\gamma}\gamma^{\alpha\beta}=\eta^{\alpha\beta})$ with $\eta^{\tau\tau}=-1,\eta^{\sigma\sigma}=1$ and $\eta^{\tau\sigma}=\eta^{\sigma\tau}=0$ to solve the equations of motion. This evidently is the main difference between the construction in  \cite{spiky1} and the one in \cite{JJ}, although they can be related via worldsheet conformal transformations.
	
	To start with, we try to reproduce \textit{Kruczenski} solution in the conformal gauge, without any flux present (i.e. $b_{MN}=0$), by taking the ansatz
	\begin{equation}\label{ansatz}
	    t=\tau + f(\sigma),~ \theta=\omega\tau+g(\sigma), ~\rho=\rho(\sigma)
	\end{equation}
	Here $\omega$ is a constant winding number, $f$ and $g$ are generic functions of $\sigma$ and the global metric of $AdS_3$ is given by
	$$ds^2=-\cosh^2{\rho}~dt^2+d\rho^2+\sinh^2\rho~ d\theta^2.$$
	Throughout this paper, we will call this string solution as \textit{Jevicki-Jin} solution.
	Now from equations of motion for $t$ and $\theta$ will give us the functional forms of $f^\prime$  and $g^\prime$as follows,
	\begin{equation}
	    f^\prime(\sigma)=\frac{a}{\cosh^2{\rho}} \hspace{8mm} g^\prime(\sigma)=\frac{b}{\sinh^2{\rho}} \ ,
	\end{equation}
	where prime $(')$ denotes the derivative with respect to $\sigma$.  The constants  $a$ and $b$ will be fixed later. On the other hand, from the first Virasoro constraint, we get a condition which can be checked to be equivalent to the equation of motion for $\rho$, hence making our solutions self-consistent.
%	\begin{equation}
%	    \rho^{\prime2}=\cosh^2\rho(1+f^{\prime2})-\sinh^2\rho(\omega^2+g^{\prime 2})
%	\end{equation}
	The second Virasoro constraint leads us to the relation between the constants,
%	\begin{equation}
%	    -\cosh^2\rho f^\prime(\sigma)+\omega\sinh^2\rho g^\prime(\sigma)=0
%	\end{equation}
%	From the second Virasoro constraint we get
	 $a=\omega b$. A consistent choice of $b$ can get us to $a$ as follows,
	\begin{equation} \label{constants}
	     b=\frac{\sinh2\rho_0}{2}; \hspace{5mm} a=\frac{\omega\sinh2\rho_0}{2}
	\end{equation}
	The constant $\rho_0$ can be chosen based on physical considerations. In fact, with these choices, we can now write the equation of motion for $\rho$ using (\ref{constants}) in the following suggestive form
	\begin{equation}
	    \rho^\prime(\sigma)=\frac{\sqrt{(\cosh^2{\rho}-\omega^2\sinh^2\rho)(\sinh^22\rho-\sinh^2 2\rho_0)}}{\sinh2\rho}.
	\end{equation}
	The integral can be done between two points, the `valley' points at $\rho=\rho_0$ where the profile reaches at it's low and the `spike' points at $\rho=\rho_1=\coth^{-1}\omega$ where the derivative $\rho^\prime$ vanishes i.e. where the cusp forms. The solution to this equation gives us the total string profile and can be given by,
		\begin{equation}
	    \rho(\sigma)=\frac{1}{2}\cosh^{-1}\left(\cosh{2\rho_1}~\textbf{cn}^2(u|k)+\cosh{2\rho_0}~\textbf{sn}^2(u|k)\right).
	\end{equation}
	Here $sn$ and $cn$ are the usual Jacobi elliptic functions and  we have defined,
	\begin{equation}
	    k=\sqrt{\frac{\cosh{2\rho_1}-\cosh{2\rho_0}}{{\cosh{2\rho_1}+\cosh{2\rho_0}}}}, \hspace{5mm}u=\sqrt{\frac{\cosh{2\rho_1}+\cosh{2\rho_0}}{{\cosh{2\rho_1}-1}}}\sigma.
	\end{equation}
	Now proceeding similarly as we did earlier i.e. by solving the equations we find $f$ and $g$ to be,
	\begin{equation}
	    f=\frac{\sqrt{2}\omega\sinh{2\rho_0}\sinh\rho_1}{(\cosh2\rho_1 +1)\sqrt{\cosh{2\rho_1}+\cosh{2\rho_0}}}~\mathbf{\Pi}\left(\frac{\cosh{2\rho_1}-\cosh{2\rho_0}}{\cosh2\rho_1 +1},x,k\right)
	\end{equation}
	\begin{equation}
	     g=\frac{\sqrt{2}\sinh{2\rho_0}\sinh\rho_1}{(\cosh2\rho_1 -1)\sqrt{\cosh{2\rho_1}+\cosh{2\rho_0}}}~\mathbf{\Pi}\left(\frac{\cosh{2\rho_1}-\cosh{2\rho_0}}{\cosh2\rho_1 -1},x,k\right).
	     \vspace{3mm}
	\end{equation}
	Here $\Pi$ is the usual incomplete Elliptic integral of the third kind. Also the variable $x$ is given by the amplitude function $x=\textbf{am}(u|k)$. Note here, spiky string solution could be obtained by lifting the minimum value of $\rho$ i.e. $\rho_0\neq 0$ as compared to GKP or folded string solution. However, one can bring this solution to that of GKP by demanding the valleys goes to zero and imposing the right boundary conditions making number of cusps on the string to be $N=2$. In the sense of boundary conditions, this string appears quite different from its GKP cousin.
	
	\subsection{String boundary conditions}
	We aim to understand the periodicity of the spiky string solution here. The solution in the conformal gauge is special in the sense of periodicity as the two time coordinate $t$ and $\tau$ are not equivalent here, as opposed to Kruczenski spiky string case. One could easily see that the derivative of string profile at the spike point does not diverge here, instead $\rho^\prime(\sigma)|_{spike}=0$ for this case. So one must be very careful in the choice of time to ensure the full string is closed.
	
	 We start with the full solution in this case, given by (2.8). One could see that the profile is implicitly function of $\sigma$ i.e. $\rho(u(\sigma))$. We now demand that the periodicity we require in $\sigma$ direction for a closed string is reflected in $u(\sigma)$ itself, so as to speak,
	$ \sigma \in [0,L] \implies u(\sigma) \in (0,\tilde L)$.
	Now, we can see that $\rho(u(\sigma))$ is naturally periodic with a non trivial (real) period of $2\textbf{K}(k)$ due to the presence of Jacobi elliptic functions. It can be verified that the associated string profile is such that it starts off at $\rho(0)=\rho_1$ at a spike, reaches a valley at $\rho(\textbf{K}(k))=\rho_0$ and again goes upto a spike at $\rho(2\textbf{K}(k))=\rho_1$, thus completing a full segment. So we can impose a periodicity of $\tilde L= 2N\textbf{K}(k)$ on the $u(\sigma)$ here, with $N$ being the total number of spikes on a given closed string. Using the expression for $u(\sigma)$, this leads us to the periodicity being,
	\begin{equation}
	    L = 2N\mathbf{K}(k)\sqrt{\frac{\cosh{2\rho_1}-1}{{\cosh{2\rho_1}+\cosh{2\rho_0}}}},
	\end{equation}
    which defines the construction of a closed spiky string. It is important to note that the `angular' functions $f(\sigma)$ and $g(\sigma)$ are not periodic with same properties, actually they are pseudo-periodic with $L' = \frac{L}{2N}$.
%    \begin{equation}
%         f({\sigma}+2mL)=f({\sigma})+\frac{\sqrt{2}\omega\sinh2\rho_0 \sinh\rho_1}{(\cosh2\rho_1 + 1)\sqrt{\cosh2\rho_1 +\cosh2\rho_0}} 2m\Pi(n_+,k)
%    \end{equation}
%    \begin{equation}
%         g({\sigma}+2mL)=g({\sigma})+\frac{\sqrt{2}\sinh2\rho_0\sinh \rho_1}{(\cosh2\rho_1-1)\sqrt{\cosh2\rho_1 +\cosh2\rho_0}} 
%         2m\Pi(n_-,k) \vspace{3mm}
%    \end{equation}
%    Where $$n_+=\frac{\cosh2\rho_1-\cosh2\rho_0}{\cosh2\rho_1+1}~~~n-=\frac{\cosh2\rho_1-\cosh2\rho_0}{\cosh2\rho_1-1}$$ \\ 
    
    We want to have a closed string at constant global time $t$, hence we need to substitute $\tau = t - f({\sigma})$ into the original ansatz 
   (\ref{ansatz}) for $\theta$, thus finding $\theta(t,{\sigma})=\omega t + g({\sigma}) - \omega f({\sigma})$. Having this, we impose a closed boundary condition $\theta(t,L)=\theta(t,0)+2n\pi$, $n\in\mathbb{Z}$ being the winding number of the string. We can easily see here that $\theta(t,L) - \theta(t,0)=2N\Delta\theta$, where the angular difference:
    \begin{eqnarray}
        \Delta\theta & = & \frac{\sqrt{2}\sinh 2\rho_0 \sinh \rho_1} {\sqrt{\cosh2\rho_1+\cosh2\rho_0}} \left[\frac{\mathbf\Pi(n_-,k)}{\cosh2\rho_1 - 1} - \frac{\omega^2\mathbf\Pi(n_+,k)}{\cosh2\rho_1+1}\right]\nonumber\\
  		& = & \frac{\sinh 2\rho_0}{\sqrt{2} \sinh \rho_1 \sqrt{\cosh2\rho_1+\cosh2\rho_1}}\left[ \mathbf\Pi(n_-,k) - \mathbf\Pi(n_+,k) \right].
	 \end{eqnarray}
	 With the definitions $$n_+=\frac{\cosh2\rho_1-\cosh2\rho_0}{\cosh2\rho_1+1}~~~n_{-}=\frac{\cosh2\rho_1-\cosh2\rho_0}{\cosh2\rho_1-1}.$$
    The closenessd condition for the string then transcends to 
$ \Delta\theta = \frac{n\pi}{N},$
    where the $N$ spikes on the strings are located at the specific positions
$      \sigma_m = 2mL,~~~~m = 0,....,N-1.$
    One can then solve the above equation fixing for the number of desired spikes and winding number, together either with position of the `valley' points $\rho_0$ or that of the `spike' points $\rho_1$ and get a set of parameters describing a perfectly closed string solution.
    
    \subsection{Dispersion relation}
    Now we find the energy and angular momentum of these configurations for completeness. The conserved Noether charges for this case are defined as, 
    \begin{equation}
        E=-2N\int\frac{\partial L} {\partial\dot{t}} d\sigma,~~~S=2N\int\frac{\partial L}{\partial\dot{\phi}}d\sigma
    \end{equation}
     When evaluated, these integrals give us the following,
%    
%    $$E=2n\int-2\cosh^2\rho\frac{d\sigma}{d\rho}d\rho$$
%    
%    solving this we get
    \begin{equation}
        E=\frac{N\sqrt{\lambda}}{2\pi}\sqrt{\frac{\cosh2\rho_1-1}{\cosh2\rho_1+\cosh2\rho_0}}[(\cosh2\rho_1+\cosh2\rho_0)\mathbf{E}(k)-2\sinh^2\rho_0~\mathbf{K}(k)]
    \end{equation}
%     Where we have multiplied 2n with E as well as for n spikes.\\
%     similarly we get the spin to be
     
%     $$S=2n\int2\omega\sinh^2\rho\frac{d\sigma}{d\rho}d\rho$$
%     
%	Solving this we get S to be
	\begin{equation}
	    S=\frac{N~\omega\sqrt{\lambda}}{2\pi}\sqrt{\frac{\cosh2\rho_1-1}{\cosh2\rho_1+\cosh2\rho_0}}\left[~(\cosh2\rho_1+\cosh2\rho_0)~\mathbf E(k)-2\cosh^2\rho_0~\mathbf K(k)\right]
	\end{equation}
	
%	Then we find the dispersion relation between E and S.
%	 \begin{equation}
%	 \begin{split}
%	    E-&S=\frac{n\sqrt{\lambda}}{2\pi}\sqrt{\frac{\cosh2\rho_1-1}{\cosh2\rho_1+\cosh2\rho_0}}\\&[(1-\omega)(\cosh2\rho_1+\cosh2\rho_0)~\mathbf E(k)+(-2\sinh^2\rho_0+2\omega\cosh^2\rho_0)~\mathbf K(k)] 
%	 \end{split}
%	 \end{equation}\vspace{0.3cm}
	 
	 Taking the limit $\rho_1>>\rho_0$ (infinite spike limit) and expanding the charges with the assumption
%	 \begin{equation}
%	     k^2=\frac{1}{\omega^2}
%	 \end{equation}
%	 \begin{equation}
%	     \sqrt{\frac{\cosh2\rho_1-1}{\cosh2\rho_1+\cosh2\rho_0}}= \frac{1}{\omega}
%	 \end{equation}
%	 \begin{equation}
%	     \cosh2\rho_1+\cosh2\rho_0=\frac{2\omega^2}{\omega^2-1}
%	 \end{equation}
%	 So
%	 \begin{equation}
%	     E-S=\frac{n\sqrt{\lambda}}{\pi}~~\left[~\mathbf K(k)-\frac{\omega}{\omega+1}~\mathbf E(k)~\right]
%	 \end{equation}
%	 \begin{equation}
%	    S=\frac{n\sqrt{\lambda}}{\pi}~\left [\frac{\omega^2}{\omega^2-1}~\mathbf E(k)-\mathbf K(k)~\right]
%	 \end{equation}
	 $\omega \to 1$ we get,
%	 \begin{equation}
%	     E-S=\frac{n\sqrt{\lambda}}{\pi}\left[~\mathbf K(k)-\frac{1}{2}\mathbf E(k)~\right]
%	 \end{equation}
%	 \begin{equation}
%	     S=\frac{n\sqrt{\lambda}}{\pi}~\left[~\frac{1}{2(\omega-1)}\mathbf E(k)-\mathbf K(k)~\right]
%	 \end{equation}\vspace{0.3cm}
%	 Now the series expansion of elliptic integral gives
%	 $$ E(k)=1$$
%	 $$K(k)=\frac{3}{2}\log2 - \frac{1}{2}\log(-1 + \omega)$$
%	 So
%	 \begin{equation}
%	     E-S= \frac{n\sqrt{\lambda}}{\pi}\left[-\frac{1}{2}+\frac{1}{2}\log8 - \frac{1}{2}\log(-1 + \omega)\right]
%	 \end{equation}
%	 \begin{equation} 
%	     S=  \frac{n\sqrt{\lambda}}{\pi}~~\left[\frac{1}{2(\omega-1)}-(~\frac{3}{2}\log2 - \frac{1}{2}\log(-1 + \omega))\right]
%	 \end{equation}\vspace{0.3cm}\\
%	 So $\log S=-\log(\omega-1)$
	 which gives the dispersion relation to be,
	\begin{equation}
	    E-S=\frac{N\sqrt{\lambda}}{\pi}\left[\frac{1}{2}\log S +..\right].
	\end{equation}
	Which reproduces the usual GKP string dispersion relation with the well-known logarithmic divergence when we put $N=2$.
%	\begin{equation}
%	    E-S= \frac{n\sqrt{\lambda}}{2\pi}~\log S+...
%	\end{equation}
	\section{N-Spike strings with fluxes}
	Now we move on to discuss the fate of the N-spike string as soon as the NS-NS flux is turned on.
	  Let us start with the AdS$_3$ metric with both R-R and NS-NS two form fields,
	\begin{equation} \label{defn}
	    ds^2_{AdS_3} = -\cosh^2\rho dt^2+d\rho^2+\sinh^2\rho d\theta^2,~~
	~~~
	    b_{t\theta}=q\sinh^2\rho,~~~c_{t\theta} = \sqrt{1-q^2}\sinh^2\rho,
	\end{equation}
	where $q$ is the flux parameter introduced before. Here since we are dealing with fundamental string, it only couples to the NS-NS flux. 
	
	We must mention here that there is a gauge freedom in the choice of the 2-form NS-NS-field, since the supergravity equations of motion instead determine the three-form field strength, $H^{(3)}=d B$. For interesting details of this gauge freedom one can look at \cite{B.Hoare}. For example, the 2-form field in $AdS_3$ can be written schematically as 
$-\frac{q}{2}(\cosh 2\rho + \chi)$, with $\chi$ being a constant that gives rise to a constant ambiguity term in the WZ part of the action. This constant ambiguity $\chi$ can be chosen via imposing  physical 
requirements on the classical string solution itself. In our case this extra constant will simply generate a constant shift in the desired  dispersion relations for the charges, and thus can be chosen suitably without any loss of generality. For our case, we will continue using the fields as written in the form of (\ref{defn}) above.

	\subsection{The solution}
	%	\begin{equation}
%	    S_{pol}= \frac{-\sqrt\lambda}{4\pi}\int d\sigma d\tau[\sqrt(-\gamma) \gamma^{\alpha\beta} d_\alpha X^M d_\beta X^Ng_{MN}-\epsilon^{\alpha\beta} d_\alpha X^M d_\beta X^N b_{MN}]
%	\end{equation}
	To find out the string solution with the presence of flux, we start with the same ansatz as in the earlier case (\ref{ansatz}). In this case the sigma model action takes the form,
%    \begin{equation}
%	\begin{split}
%	    S_{pol}= \frac{\sqrt\lambda}{4\pi}\int d\sigma d\tau & [-\cosh^2\rho(\dot t^2-t^{\prime2}) -\rho^{\prime2}-\sinh^2\rho(\theta^{\prime2}-\dot\theta^2)+ \\ & 2\theta^\prime \dot t q\sinh^2\rho-2\dot\theta t^{\prime} q \sinh^2\rho]
%	    \end{split}
%	\end{equation}
	\begin{equation}
	\begin{split}
	   S_{pol}= \frac{\sqrt\lambda}{4\pi} \int d\sigma d\tau & \bigg[-\cosh^2\rho(1-f^{\prime2})-\rho^{\prime2}-\sinh^2\rho(g^{\prime2}-\omega^2)\\ &+ 2g^\prime q\sinh^2\rho-2\omega q f^\prime \sinh^2\rho \bigg]
	\end{split}
	\end{equation}
	 As before, we can solve the equation of motion for $t$ and $\theta$, which boils down to equations for $f$ and $g$,
	 \begin{equation}
	      f^\prime(\sigma)=\frac{\bar a+\omega q\sinh^2\rho}{\cosh^2\rho},\hspace{2cm} g^\prime(\sigma)=\frac{q\sinh^2\rho+\bar b}{\sinh^2\rho}
	 \end{equation}
	Here $\bar a,\bar b$  are again constants, and using the Virasoro constraints one could again take $\bar b=\frac{\sinh2\rho_0}{2}$ and $ \bar a=\frac{\omega\sinh2\rho_0}{2}$
	as a consistent choice.
	Using this result, $f^\prime$ and $ g^\prime$ can be written as
	 \begin{equation}\label{fgflux}
	     f^\prime=\frac{\omega\sinh2\rho_0+2\omega q\sinh^2\rho}{2\cosh^2\rho},
	~~~~
	     g^\prime=\frac{\sinh2\rho_0+2q\sinh^2\rho}{2\sinh^2\rho}.
	 \end{equation}
	 The $\rho$ equation of motion, consistent with the first Virasoro constraint, turns out to be,
     $$\rho^\prime(\sigma)=\frac{\sqrt{(\cosh^2\rho-\omega^2\sinh^2\rho)(\sinh^22\rho-(\sinh2\rho_0+2q\sinh^2\rho)^2)}}{\sinh2\rho}.$$\label{rhoflux}
     Some comments are in order here. One could see from the equation that the positions of spikes in the string profile remains same i.e. $\rho_1 = \coth^{-1}\omega$, but the position of `valleys' have changed considerably due to inclusion of the flux. 
     Solving the above differential equation requires to fix the right limits of integration. The lower limit of integration in this case is given by the solutions of $\rho^\prime(\sigma)=0$, out of which we choose the particular solution,
     \begin{equation}\sinh\rho_{min}=\frac{-(1-q\sinh2\rho_0)+\sqrt{\cosh^22\rho_0-2q\sinh2\rho_0}}{2(1-q^2)}\end{equation}\label{rootq}
     Note here when we put $q=0$, the minimum value becomes $\rho_{min} = \rho_0$, exactly like in the case of no flux, and hence our choice is justified. One should also note that for $0\leq q<1$, and $\rho_0>0$, we will always have $\rho_{min}\geq \rho_0$, i.e. the new `valley' will always occur at a higher radial value than in the no-flux case. In other words the string will get `fatter' as we increase the flux. Now, integrating the differential equation, we explicitly find the profile for the spiky string, 
     \begin{equation}\label{proflux}
     \begin{split}
          \rho(\sigma)=\frac{1}{2} & \cosh^{-1}\Big(\frac{1}{(1-q^2)} [(1-q^2)\cosh2\rho_1~\textbf{cn}^2(u | k)+\\& (q\sinh2\rho_0-q^2
           +\sqrt{\cosh^22\rho_0-2q\sinh2\rho_0})~\textbf{sn}^2(u| k)]\Big),
     \end{split}
     \end{equation}
     again in terms of Jacobi functions. Here the arguments of the elliptic function are affected by the presence of flux and they read,
     \begin{equation}
         u=\sqrt{\frac{2(1-q^2)\sinh^2\rho_1+1-q\sinh2\rho_0+\sqrt{\cosh^22\rho_0-2q\sinh2\rho_0}}{2\sinh^2\rho_1}}\sigma
    \end{equation}
    and
    \begin{equation}
        k=\sqrt{\frac{(1-q^2)\cosh2\rho_1+q^2-q\sinh2\rho_0-\sqrt{\cosh^22\rho_0-2q\sinh2\rho_0}}{(1-q^2)\cosh2\rho_1+q^2-q\sinh2\rho_0+\sqrt{\cosh^22\rho_0-2q\sinh2\rho_0}}}.
    \end{equation}
     Now integrating (\ref{fgflux}) we can write the functions $f$ and $g$ to be of the following form,
    \begin{equation}
        \begin{split}
         f=\omega & \sqrt{\frac{(\cosh2\rho_1-1)}{(1-q^2)\cosh2\rho_1+q^2-q\sinh2\rho_0+\sqrt{\cosh^22\rho_0-2q\sinh2\rho_0}}}\\
         & \hspace{3cm}
         \Bigg[\frac{\sinh2\rho_0-2q}{\cosh2\rho_1+1}~\mathbf{\Pi}(n_f,x,k)+q \mathbf{K}(x,k)\Bigg]
    \end{split}
    \end{equation}\vspace{0.3cm}
    \begin{equation}
    \begin{split}
        g=&\sqrt{\frac{(\cosh2\rho_1-1)}{(1-q^2)\cosh2\rho_1+q^2-q\sinh2\rho_0+\sqrt{\cosh^22\rho_0-2q\sinh2\rho_0}}}\\
        &\hspace{3.5cm}\Bigg[\frac{\sinh2\rho_0}{\cosh2\rho_1-1}~\mathbf{\Pi}(n_g,x,k)+q \mathbf{K}(x,k)\Bigg].
    \end{split}
    \end{equation}
    
         \begin{figure}[t]
{\includegraphics[scale=0.6]{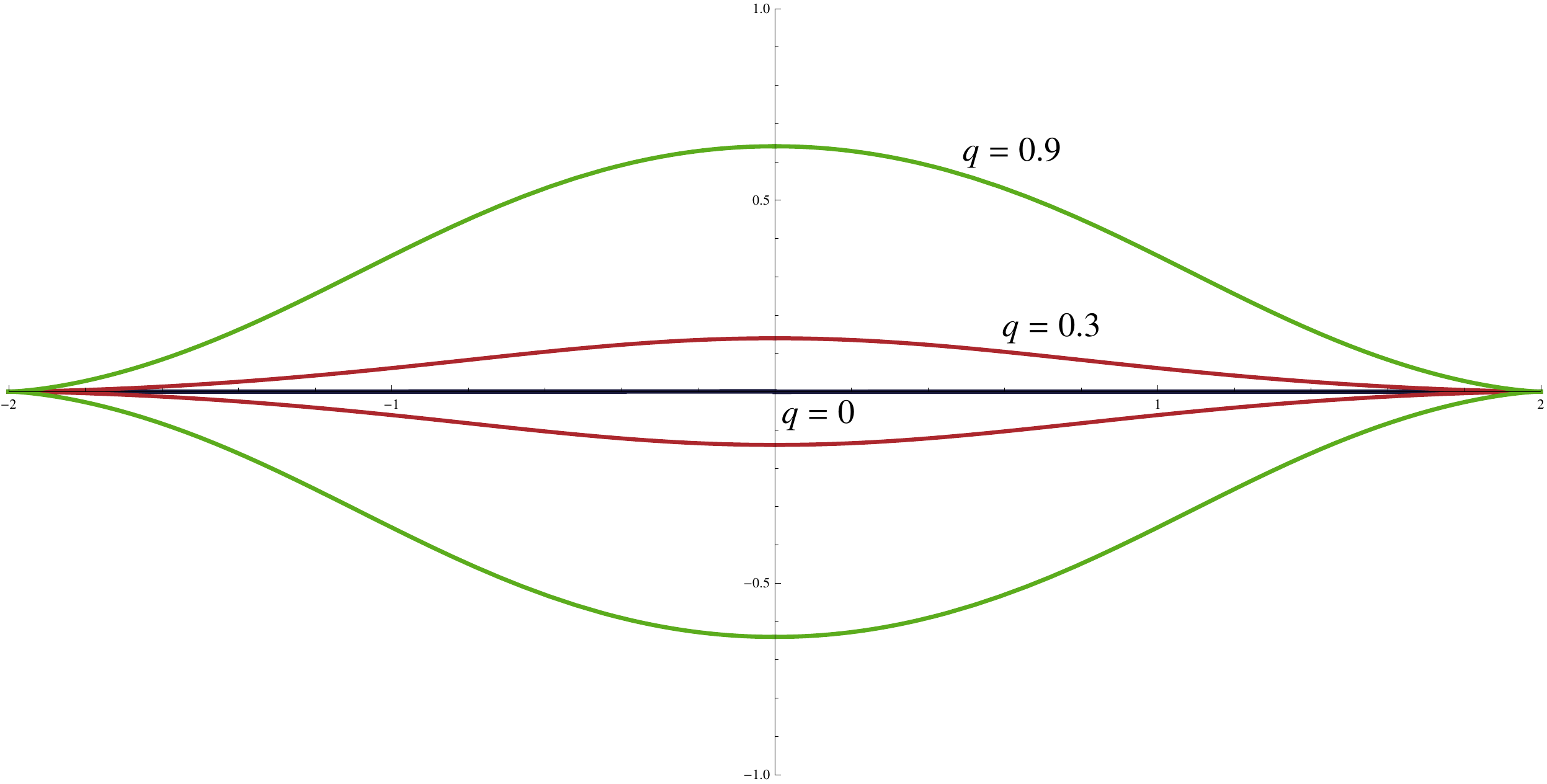}}
\caption{The $N=2$ i.e. ``folded'' version of the spiky string is plotted for different values of $q$. One must immediately note that for $q>0$ the string does not pass through $\rho=0$ point, which is consistent with our computations. As $q\to 1$, the string gets more and more `fat'.}\label{fig1}
\end{figure}

   As in the earlier case, $x$ is given by the amplitude function $\mathbf{am}(u|k)$, which again is flux dependent now. Here the parameters $n_f$ and $n_g$ are given as,
    \begin{equation}
        n_f=\frac{(1-q^2)\cosh2\rho_1+q^2-q\sinh2\rho_0-\sqrt{\cosh^22\rho_0-2q\sinh2\rho_0}}{(1-q^2)(\cosh2\rho_1+1)}
    \end{equation}\vspace{0.3cm}
    \begin{equation}
        n_g=\frac{(1-q^2)\cosh2\rho_1+q^2-q\sinh2\rho_0-\sqrt{\cosh^22\rho_0-2q\sinh2\rho_0}}{(1-q^2)(\cosh2\rho_1-1)}.
    \end{equation}
%    Where $x=am(u,k)$ and $\Pi(n,x,k)$ is the incomplete elliptic integral.
    With this solution at hand, we move on to define proper boundary condition for the string in the presence of flux.
    
\begin{figure}[t]
{\includegraphics[scale=0.6]{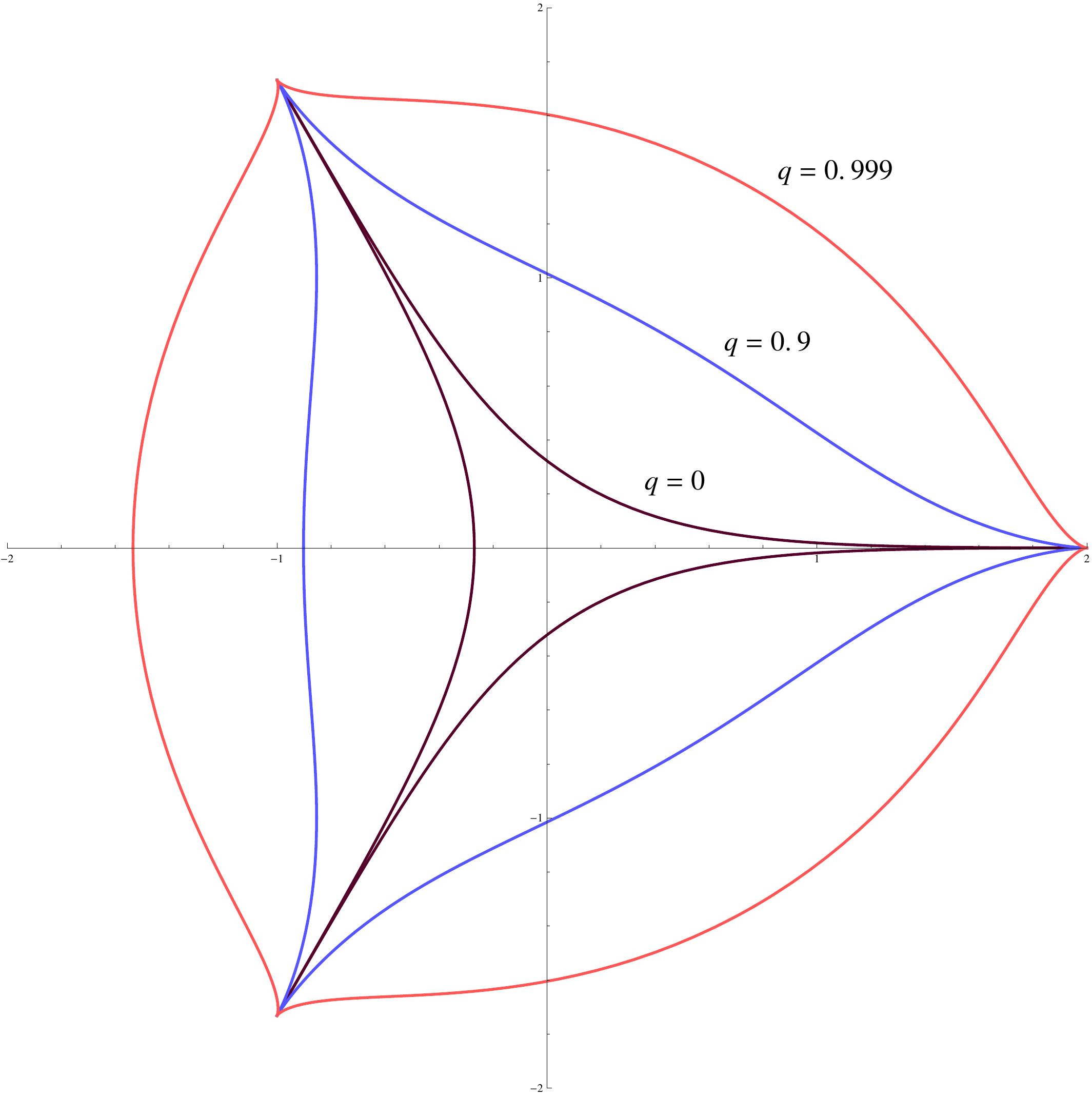}}
\caption{The $N=3$ string is plotted for different values of $q$. As one increases $q$ the string becomes more `fat'. Near $q\to 1$ regime, the `fatness' increases very fast as the valley positions go to infinity rapidly. }\label{fig2}
\end{figure}

    For completeness, we plot the string profiles in figure \ref{fig1} and figure \ref{fig2}, for $N=2$ and $N=3$ respectively, for different values of flux parameters turned on. One can see that in $N=2$ case, the string stops passing through the centre of $AdS$ ($\rho=0$) as the flux is cranked up from zero. This is somewhat expected from our analysis, but also very intriguing. Since the $N=2$ string is supposed to be the analogue to the folded sting, and by definition it should be a \textit{straight} string. But it seems with the inclusion of the flux, when we impose the constraint of being a closed string, it simply changes the usual folded string boundary conditions. One can speculate here that this is why the boundary condition used in \cite{Rotating1} gives a complicated answer, as the authors assumed even with the fluxes the folded string should pass through the centre of $AdS$. In that sense our construction describes a somewhat different object than the usual idea of a folded string. For the $N=3$ case we can see clearly how the string becomes `fat' with inclusion of the flux. Near $q\to 1$, it almost becomes a circular string with the cusp points still present as the valley positions rapidly increase towards the $AdS$ boundary too.
    
    \subsection{Modified String Boundary conditions}
    As we can easily anticipate, there should be modification in the periodicity of the spiky string solution in the presence of the NS-NS  flux. Starting with the full solution in the mixed flux case given by (\ref{proflux}), we see again the profile is implicitly function of $\sigma$ i.e $\rho(u(\sigma))$ as was the case with the string solution without flux. Now we demand that the periodicity we require in $\sigma$ direction for a closed string again is reflected in $u(\sigma)$ itself, i.e. $       \sigma \in [0,L] \implies u(\sigma) \in[0,\tilde L].$ Here the periodicity of $u(\sigma)$ is $\tilde{L}$ which is $2N\mathbf{K}(k)$.
%     Now it is clear from the periodicity of elliptic functions that $\rho(u(\sigma))$ is naturally periodic with a non trivial (real) period of 2$\textbf K(k)$ . Note that here the string profile starts off at $\rho(0)=\rho_1$ at a spike, reaches a valley at $\rho(\textbf K(k))=\rho_min$ which is not equal to $\rho_0$ here and again goes up to a spike at  $\rho(\textbf K(k))=\rho_1$, thus completing one full segment.So we can impose a periodicity of $\tilde L =2N\textbf K(k)$ on the $u(\sigma)$ here, with $N$ being the no. of spikes.
     So now using the expression for $u(\sigma)$, we can find the period for the spiky string as,
     \begin{equation}
         L=2N\mathbf{K}(k)\frac{\sqrt{\cosh{2\rho_1}-1}}{\sqrt{{2(1-q^2)\sinh^2\rho_1+1-q\sinh2 \rho_0+\sqrt{\cosh^22\rho_0-2q\sinh2\rho_0}}}},
     \end{equation}  
     which is crucial for the construction of a closed spiky string with $N$ segments.

    Similarly as before, we can now impose $\theta(t,L)=\theta(t,0)+2n\pi$, for $n\in\mathbb{Z}$ being the winding number. To close the string we must subsequently have a closed solution at constant global time $t$, implying $\theta(t,L) - \theta(t,0)=2N\Delta\theta$, where:
     \begin{equation}
       \begin{split}
           \Delta\theta = &\frac{1}{\sqrt{2} \sinh \rho_1 \sqrt{{2(1-q^2)\sinh^2\rho_1+1-q\sinh2 \rho_0+ \sqrt{\cosh^22\rho_0-2q\sinh2\rho_0}}}}\\&\Bigg[\sinh2\rho_0 ~\mathbf\Pi(n_g,k) - (\sinh2\rho_0-2q)~\mathbf\Pi(n_f,k) -2q~\mathbf K(k) \Bigg]
       \end{split}
     \end{equation}
     Here, $n_f$ and $n_g$ has been defined as before.
     The closeness condition boils down to the condition
$     \Delta\theta = \frac{n\pi}{N},$
     where the $N$ spikes on the string are located symmetrically at the positions defined by $\rho_1$.
     One can then solve the above equation fixing for the number of desired spikes and winding number, together either with position of the `valley' points $\rho_{min}$ or that of the `spike' points $\rho_1$ and once again get a perfectly closed string solution.

     \subsection{Dispersion Relation}
   Once we have the string profile ready, we can move on to the focus of the paper, finding the dispersion relation among conserved charges for the string. We write down the energy and angular momentum of these configuration for completeness,
    \begin{equation}
        E=-2N\int\frac{\partial L} {\partial\dot{t}}d\sigma, ~~~S=2N\int\frac{\partial L}{\partial\dot{\theta}}d\sigma
    \end{equation}
    Using (\ref{rhoflux}) we can transform the Noether charges as integrals on $\rho$,
    \begin{eqnarray}
        E&=&N\frac{\sqrt{\lambda}}{\pi}\int\frac{(\cosh^2\rho-(q\sinh2\rho_0)/2-q^2\sinh^2\rho)\sinh2\rho}{\sqrt{(\cosh^2\rho-\omega^2\sinh^2\rho)(\sinh^22\rho-(\sinh2\rho_0+2q\sinh^2\rho)^2)}}d\rho, \nonumber \\
  S&=&N\frac{\omega\sqrt{\lambda}}{\pi}\int\frac{(\sinh^2\rho-q\sinh^2 \rho(\frac{\sinh2\rho_0+2q\sinh^2\rho}{2\cosh^2\rho}))\sinh2\rho} {{\sqrt{(\cosh^2\rho-\omega^2\sinh^2\rho)(\sinh^22\rho-(\sinh2\rho_0+2q\sinh^2\rho)^2)}}}d\rho. \nonumber
    \end{eqnarray}

       %  Taking $\sinh^2\rho= t  \implies \sinh2{\rho} d\rho= dt $
   % \begin{equation}
    %    \begin{split}
    %     &E=n\frac{\sqrt{\lambda}(1-q^2)}{\pi\sqrt{1-q^2}} \int\frac{t\sqrt{-a}}{2\sqrt{(t-a)(t-b)(t-c)}}dt\\
    %     &\hspace{2cm}+\frac{\sqrt{\lambda}(1-q\sinh2\rho_0/2)}{2\pi\sqrt{1-q^2}}\int\frac{\sqrt{-a}}{2\sqrt{(t-a)(t-b)(t-c)}}dt
  %  \end{split}
  %  \end{equation}\vspace{0.6cm}
    We can integrate the charges from $\rho_{min}$ to $\rho_1$ to get closed form expressions, which we don't write explicitly here for brevity.
%    \begin{equation}
%    \begin{split}
%        E=&N\frac{\sqrt{\lambda}}{2\pi}\frac{\sqrt{(\cosh2\rho_1-1)}}{\sqrt{(1-q^2)\cosh2\rho_1+q^2-q\sinh2\rho_0+\sqrt{\cosh^22\rho_0-2q\sinh2\rho_0}}}\\ &\Bigg[\left((1-q^2)\cosh2\rho_1+q^2-q\sinh2\rho_0+\sqrt{\cosh^22\rho_0-2q\sinh2\rho_0}\right)\mathbf{E}(k)\\ & \hspace{2cm}
%        +\left(1-\sqrt{\cosh^22\rho_0-2q\sinh2\rho_0}\right)\mathbf{K}(k)\Bigg]
%    \end{split}
%    \end{equation}
%    Similarly we get from the spin integral,
%       % Taking $\sinh^2\rho= t  \implies \sinh2{\rho} d\rho= dt $
%   % \begin{equation}
%   %     S=n\frac{\omega\sqrt{\lambda}\sqrt{-a}}{\pi} \int\frac{t-\frac{qt\sinh2\rho_0}{2(1+t)}-\frac{q^2t^2}{1+t}}{2\sqrt{(t-a)(t-b)(t-c)(1-q^2)}}dt\vspace{0.6cm}
%  %  \end{equation}\vspace{0.6cm}
%%  On solving which gives
%    \begin{equation}
%     \begin{split}
%        S=N\frac{\omega\sqrt{\lambda}}{2\pi}\frac{\sqrt{(\cosh2\rho_1-1)}}{\sqrt{(1-q^2)\cosh2\rho_1+q^2-q\sinh2\rho_0+\sqrt{\cosh^22\rho_0-2q\sinh2\rho_0}}}\\
%       \Bigg[\left((1-q^2)\cosh2\rho_1+q^2-q\sinh2\rho_0+\sqrt{\cosh^22\rho_0-2q\sinh2\rho_0}\right)\mathbf{E}(k)\\
%        +\left(2q^2-1-\sqrt{\cosh^22\rho_0-2q\sinh2\rho_0}\right)\mathbf{K}(k)+2\left (\frac{q\sinh2\rho_0-2q^2}{\cosh2\rho_1+1}\right)\mathbf{\Pi}(n_g,k)\Bigg]
%     \end{split}
%    \end{equation}
%%    Where $$n=\frac{(1-q^2)\cosh2\rho_1+q^2-q\sinh2\rho_0+\sqrt{\cosh^22\rho_0-2q\sinh2\rho_0}}{(1-q^2)(\cosh2\rho_1-1)}$$\vspace{0.6cm}
%Here $n_g$ has been defined as before. 
However, to find the desired dispersion relation, one has to start from the combination of charges $E-S$, where
    \begin{equation}
    \begin{split}
        E-S=& N\frac{\sqrt{\lambda}}{2\pi}\frac{\sqrt{(\cosh2\rho_1-1)}}{\sqrt{(1-q^2)\cosh2\rho_1+q^2-q\sinh2\rho_0+\sqrt{\cosh^22\rho_0-2q\sinh2\rho_0}}}\\ &\Bigg[(1-\omega)\left((1-q^2)\cosh2\rho_1+q^2-q\sinh2\rho_0+\sqrt{\cosh^22\rho_0-2q\sinh2\rho_0}\right)\mathbf{E}(k)\\ &
         +\left((1-\omega)\left(1-\sqrt{\cosh^22\rho_0-2q\sinh2\rho_0}\right)+2\omega(1-q^2)\right) \mathbf{K}(k) \\& -2\omega\left(\frac{q\sinh2\rho_0-2q^2}{\cosh2\rho_1+1}\right)\mathbf{\Pi}(n_g,k)\Bigg].
         \end{split}
    \end{equation}
    %Here $n_g$ has been defined as before.
    Now, our objective is to go to the large spin limit of the expression, which can be reinterpreted as the limit $\rho_1>>\rho_0$, i.e. when the spikes become infinitely long in comparison to the position of the valleys. Precisely in that case the cusps will try to reach the $AdS$ boundary and we would be able to effectively compute the boundary theory observable from this bulk string. In this limit, we find the
    arguments of our elliptic functions behave in the following way,
    \begin{equation}
        k^2=\frac{1-q^2}{\omega^2-q^2}, \hspace{1cm}
        n_g=\frac{1}{\omega^2},
    \end{equation}
    So our expression for the difference in charges can be written in the  form,
    \begin{equation}
        E-S=N\frac{\sqrt{\lambda}}{\pi}\frac{1}{\sqrt{\omega^2-q^2}}\left[\omega (1-q^2)\mathbf{K}(k)+\frac{q^2(\omega^2-1)}{\omega}\mathbf{\Pi}(n_g,k)-\frac{\omega^2-q^2} {1+\omega}\mathbf{E}(k)\right]
    \end{equation}
    and similarly the integral for spin can be evaluated in the limit,
    \begin{equation}
        S=N\frac{\omega\sqrt{\lambda}}{\pi}\frac{1}{\sqrt{\omega^2-q^2}}\left[\frac{\omega^2-q^2}{\omega^2-1}\mathbf{E}(k)-(1-q^2)\mathbf{K}(k)+\frac{q^2(1-\omega^2)}{ \omega^2}\mathbf{\Pi}(n_g,k)\right].
    \end{equation}
     Now we should bear in mind since even in the flux case $\rho_1 = \coth^{-1}\omega$, one should also take $\omega\to 1$ to find consistent dispersion relations in the large spin limit. Imposing this limit and using the expansions of the elliptic functions in this regime,
%    \begin{equation}
%        E-S=n\frac{\sqrt{\lambda}}{\pi}\sqrt{1-q^2}\left[K(k)-\frac{1}{2}E(k)\right]
%    \end{equation}
%    and
%    \begin{equation}
%        S=n\frac{\sqrt{\lambda}}{\pi}\sqrt{1-q^2}\left[\frac{1}{2(\omega-1)}E(k)-K(k)\right]
%    \end{equation}
  $$ \mathbf{E}(k)\approx1,~~~~ \mathbf{K}(k)\approx\frac{1}{2}\log(16(1-q^2))-\frac{1}{2}\log(\omega-1),$$
    we can expand our conserved charges in the following way
    \begin{equation}
        S=N\frac{\sqrt{\lambda}}{\pi}\sqrt{1-q^2}~\left[\frac{1}{\omega-1}-\frac{1}{2}\log (16(1-q^2))+...\right]
    \end{equation}
   % which gives $\log S=-\log (\omega-1)$
    and the difference between the charges become,
    \begin{equation}
        E-S=N\frac{\sqrt{\lambda}}{\pi}\sqrt{1-q^2}~\left[\frac{1}{2}\log (16(1-q^2))-\frac{1}{2}\log (\omega-1)-\frac{1}{2}+...\right].
    \end{equation}
    Collecting the above expressions, we can finally write the dispersion relation for the string,
    \begin{equation}\label{main}
         E-S=N\frac{\sqrt{\lambda}}{\pi}\sqrt{1-q^2}~\left[\frac{1}{2}\log S+\mathcal{O}\left(\frac{1}{S}\right)+...\right].
    \end{equation}
    The above relation is exact in $q$, and can be crosschecked by looking at the result when $q=0$ which gives the pure RR result of the usual spiky string. It seems the dispersion relation retains its $\log S$ behaviour at the leading order, but the string tension has been scaled by the factor $\sqrt{1-q^2}$. This is analogous to the result obtained in \cite{stepanchuk} using a worldsheet transformation on the $q=0$ folded string ($N=2$) moving in $AdS_3\times S^1$. As we mentioned before, in \cite{multispike} the dispersion relation for the spiky string with fluxes was obtained perturbatively in $q$, and the small $q$ expansion of the above result agrees with it, providing a validation for us. 
    %    \begin{equation}
%        E-S=n\frac{\sqrt{\lambda}}{2\pi}\log S+...
%    \end{equation}
    \subsection{Case of $q=1$: Pure NS-NS flux}
    Let us consider the case where the background $AdS_3$ is supported purely by NS-NS flux and absence of  any R-R flux. This is a special case and can't be naively studied by taking $q=1$ on all the results obtained in previous sections. Instead, one has to start from the first principles, put $q=1$ and re-analyze the problem.

         \begin{figure}[t]
         \begin{centering}
{\includegraphics[scale=0.5]{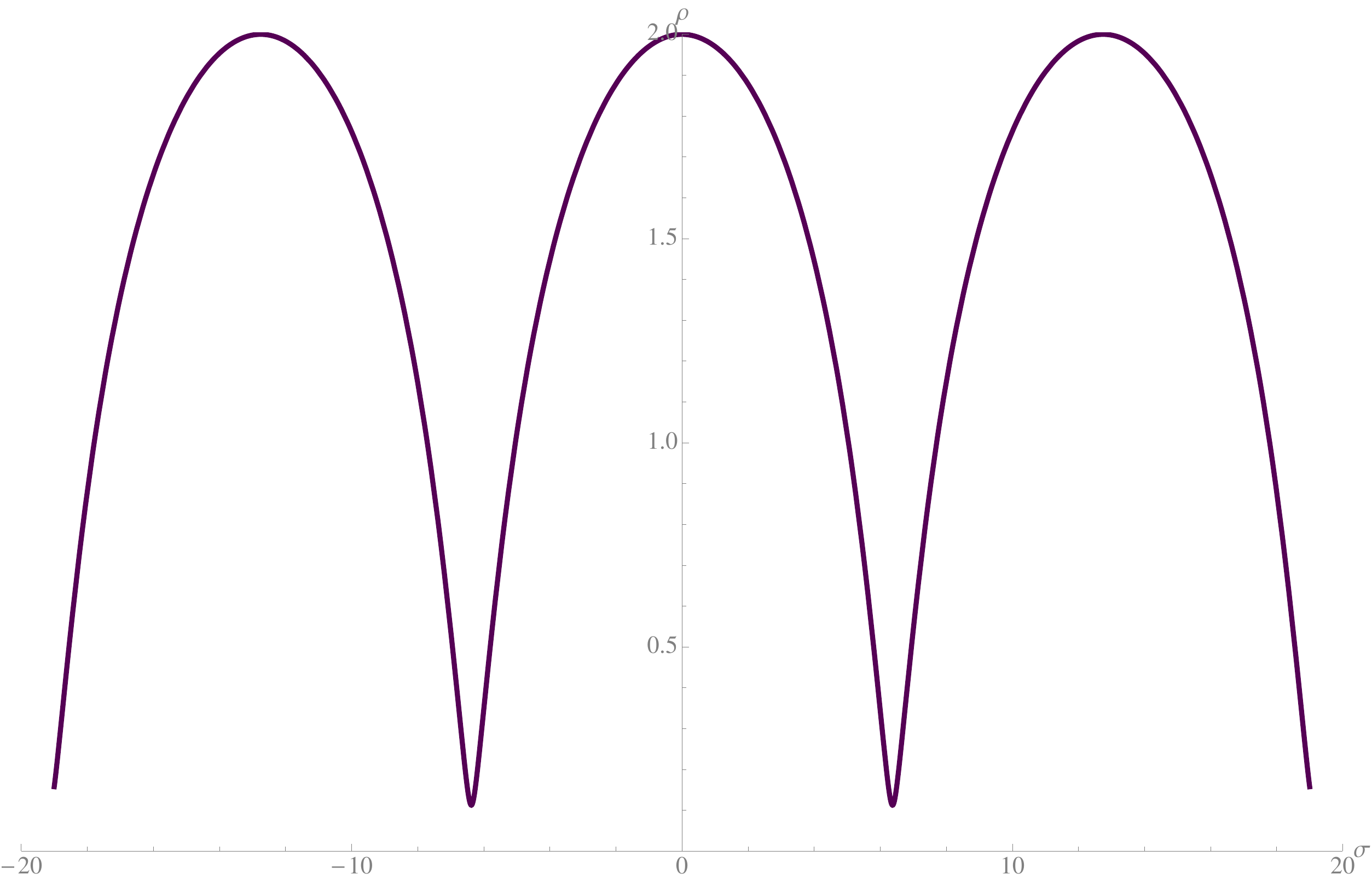}}
\caption{The string profile $\rho(\sigma)$ plotted against $\sigma$ for $q=1$, i.e pure NS flux case.}\label{fig3}
\end{centering}
\end{figure}
    In this limit, the string action becomes
    \begin{equation}
        \begin{split}
	    S_{pol}= \frac{\sqrt\lambda}{4\pi}\int d\sigma d\tau & \bigg[-\cosh^2\rho(\dot t^2-t^{\prime2}) -\rho^{\prime2}-\sinh^2\rho(\theta^{\prime2}-\dot\theta^2)+ \\ & 2\theta^\prime \dot t \sinh^2\rho-2\dot\theta t^{\prime}  \sinh^2\rho\bigg].
	    \end{split}
    \end{equation}
    The consistent equation of motion for  $\rho$ turns out to be the $q=1$ version of (\ref{rhoflux}),
     $$\rho^\prime(\sigma)=\frac{\sqrt{(\cosh^2\rho-\omega^2\sinh^2\rho)(\sinh^22\rho-(\sinh2\rho_0+2\sinh^2\rho)^2)}}{\sinh2\rho}.$$
    %\begin{equation}
      % \int d\sigma=\int{\frac{\sinh2\rho}{\sqrt{(\cosh^2\rho-\omega^2\sinh^2\rho)(\sinh^22\rho-(\sinh2\rho_0+2\sinh^2\rho)^2)}}} d\rho
    %\end{equation}
    %Taking $\sinh^2\rho=t$ we get
    %\begin{equation}
      % \int  d\sigma=\int{\frac{dt}{2\sqrt{(1-\omega^2)(t-a)(t-b)(1-\sinh2\rho_0)}}}
    %\end{equation}
    %where $a=\sinh^2\rho_1$ and $b=\frac{\sinh^22\rho_0}{4(1-\sinh^22\rho_0)}$
    %\begin{equation}
    %   =\frac{\sinh\rho_1}{2\sqrt{\sinh2\rho_0-1}}\int \frac{dy}{y(1+y)}
    %\end{equation}
    % Where we have taken $t-a=x$ and $\frac{x}{a-b}=y$\\
    % \begin{equation}
    %    =\frac{\sinh\rho_1}{2\sqrt{\sinh2\rho_0-1}} 2\sinh^{-1}\sqrt{y}
    % \end{equation}
   % So
   %  \begin{equation}
   %     \frac{\sqrt{\sinh2\rho_0-1}}{\sinh\rho_1}\sigma=\sinh^{-1}\sqrt{y}
   %  \end{equation}
     Looking at the equation, we again observe the cusps still occur at the same points i.e. $\rho_1 = \coth^{-1}\omega$, but the minimum value of $\rho$ has changed considerably here. The minima of the profile now lies at a changed value,
     \begin{equation}
     \sinh\rho_{min} = \frac{\sinh\rho_0 \cosh\rho_0}{\sqrt{1-\sinh 2\rho_0}},
          \end{equation}
          which is not continuously connected to the intermediate root (\ref{rootq}). We note here that the position of the valleys here increase very fast with increasing $\rho_0$, i.e. the valleys also try to reach infinite radius as well. One could think this as a situation where the string tries to ``stick'' to the boundary, as seems evident from the figure \ref{fig2}. Integrating the equation of motion from $\rho_{min}$ to $\rho_1$, we get the string profile as follows,

     \begin{equation}
         \rho(\sigma)=\frac{1}{2}\cosh^{-1}\left(\cosh2\rho_1\cosh^2u-\frac{2(1-\sinh2\rho_0)+\sinh^22\rho_0}{2(1-\sinh2\rho_0)}\sinh^2u\right)
     \end{equation}
     where we have defined a new variable, $$u=\frac{\sqrt{\sinh2\rho_0-1}}{\sinh\rho_1}\sigma.$$
     The profile even in this limit is perfectly periodic in $\sigma$ as evident from figure \ref{fig3}.
     Following the last section,  we can easily integrate the other two equations of motion to find out the expressions for $f(\sigma)$ and $g(\sigma)$, which read,
     \begin{eqnarray}
         f&=&\frac{-\omega\pi}{2}\frac{\sinh\rho_1}{\sqrt{1-\sinh2\rho_0}}\left(1+\frac{\sqrt{1-\sinh2\rho_0}}{\omega\sinh\rho_1}\right); \nonumber\\
    g&=&\frac{-\pi}{2}\frac{\sinh\rho_1}{\sqrt{1-\sinh2\rho_0}}\left(1+\frac{\sqrt{1-\sinh2\rho_0}}{\sinh\rho_1}\right).
     \end{eqnarray}
   It is also straightforward to find out the conserved charges associated to the solution,
    \begin{equation}
        E=-2N\int\frac{\partial L} {\partial\dot{t}}d\sigma, ~~~S=2N\int\frac{\partial L}{\partial\dot{\theta}}d\sigma.
    \end{equation}
    Using the $\rho$ equation of motion, we could write the energy in the integral form,
    \begin{equation}
         E=N\frac{\sqrt{\lambda}}{\pi}\int\frac{(\cosh^2\rho-(\sinh2\rho_0)/2-\sinh^2\rho)\sinh2\rho}{\sqrt{(\cosh^2\rho-\omega^2\sinh^2\rho)(\sinh^22\rho-(\sinh2\rho_0+2\sinh^2\rho)^2)}}d\rho \ ,
    \end{equation}
  %  Taking $\sinh^2\rho= t  \implies \sinh2{\rho} d\rho= dt $
  %  \begin{equation}
       %  E=n\frac{\sqrt{\lambda}}{\pi}(1-\frac{\sinh2\rho_0}{2})\frac{\sqrt{-a}}{2\sqrt{(1-\sinh2\rho_0)}}\int\frac{1}{\sqrt{(t-a)(t-b)}} dt
    %\end{equation}
   % where $$a=\frac{1}{\omega^2-1}~~~ %b=\frac{\sinh^22\rho_0}{4(1-\sinh2\rho_0)}$$
   which turns out to be very simple algebraic expression when evaluated, i.e. 
    \begin{equation}
        E=N\frac{\sqrt{\lambda}}{4}\frac{(\sinh2\rho_0-2)}{\sqrt{(1-\sinh2\rho_0)}}\sqrt{\frac{1}{\omega^2-1}}.
    \end{equation}
  
    Proceeding in the same way, we go on to obtain Spin $S$,
    \begin{equation}
         S=N\frac{\omega\sqrt{\lambda}}{\pi}\int\frac{(\sinh^2\rho-\sinh^2 \rho(\frac{\sinh2\rho_0+2\sinh^2\rho}{2\cosh^2\rho}))\sinh2\rho} {{\sqrt{(\cosh^2\rho-\omega^2\sinh^2\rho)(\sinh^22\rho-(\sinh2\rho_0+2\sinh^2\rho)^2)}}}d\rho \ ,
    \end{equation}
   % \begin{equation}
   % \begin{split}
   %     S=2n\omega\frac{\sqrt{\lambda}}{4\pi}&(1-\frac{\sinh2\rho_0}{2})\frac{1}{\sqrt{(1-\sinh2\rho_0)}} \sqrt{\frac{1}{1-\omega^2}}\\&\left[\int\frac{dt}{\sqrt{(t-a)(t-b)}} -\int\frac{dt}{(1+t)\sqrt{(t-a)(t-b)}}\right]
   %     \end{split}
   % \end{equation}
   which evaluates to similar algebraic expressions,
    \begin{equation}
        S=N\frac{\sqrt{\lambda}}{4}\frac{\sinh2\rho_0-2}{\sqrt{(1-\sinh2\rho_0)}} \left(\frac{\omega}{\sqrt{\omega^2-1}}-\frac{ 2\sqrt{1-\sinh2\rho_0}}{(\sinh2\rho_0-2)}\right).
    \end{equation}
    Collecting all the equations we get the relevant results,
    \begin{equation}
        E-S=N\frac{\sqrt{\lambda}}{4}\frac{\sinh2\rho_0-2}{\sqrt{(1-\sinh2\rho_0)}}  \left(\frac{1-\omega}{\sqrt{\omega^2-1}}+\frac{ 2\sqrt{1-\sinh2\rho_0}}{ (\sinh2\rho_0-2)}\right).
    \end{equation} 
    Now again we would like to take the large spin limit on the solution. We need to impose $\rho_1 \gg \rho_0$ which clearly implies $\omega \to 1$. Under this assumpition the expression for $E-S$ becomes,
    \begin{equation}
        E-S=N\frac{\sqrt{\lambda}}{2}.
    \end{equation}
    Note that the right hand side of the dispersion relation gives a mere constant here. However, this trivialization is not entirely unexpected. Since the $q=1$ case corresponds to the WZW limit, one should get a dispersion relation analogous to that of a `massless' case, which also appears in the case of dyonic giant magnons \cite{B.Hoare} of the theory in this limit.
    
  %  \section{Connecting static and conformal gauges}
    \section{Conclusions and Outlook}
    In this paper, we have studied N-spike strings in the conformal gauge and found out the string profile, conserved charges and large-charge dispersion relations for the string. We observed that as one increases the strength of the background flux, the ``valleys'' of the string become larger and the string becomes more and more ``fat''. The main result of the paper is simply (\ref{main}), where we present the energy-spin dispersion relation for the string, which should be indicative of anomalous dimensions of the twist operators in a dual field theory. The $N=2$ version of our result would correspond to analogues of folded strings in this spacetime, which in our case is a perfectly closed string in contrast to the case in \cite{Rotating1} where one needed a large number of windings to achieve closedness.  We also discussed the case of pure NS-NS fluxes, where the theory will flow to that of a WZW model, and found the dynamics of the string. 
    
    Now, we can hope that this calculation resolves the issue for dispersion relation of such strings. Going forward, there are more problems that one can study alongside this. The first modest follow-up could be to add an angular momentum along a $S^1$ direction and explicitly prove the results of \cite{stepanchuk}. Actually it was shown in \cite{Ishizeki:2008tx} that adding an angular momenta to a N-spike string results in the smoothening of the spikes, i.e. cusps become rounded off. It will be interesting to see if such a phenomenon occurs in the presence of NS-NS fluxes too. Similar arguments should also hold for minimal surfaces in \cite{Hernandez:2018lvh} with added non-vanishing angular momentum, and one could study the effect of fluxes on them.
    
    Another issue is that of pure NS-NS limit, where our solutions and dispersion relations become very simple. One can hope that this in conjunction with recently uncovered spin-chain picture for the WZW case \cite{purens2} sheds some light on this limit. However, one needs to have better understanding of these solutions to proceed along those lines. We would be coming back to some of these problems in future correspondences. 
    
\section*{Acknowledgements} 
We would like to thank Manoranjan Samal for useful discussions. 
Aritra Banerjee is supported in part by the Chinese Academy of Sciences (CAS) Hundred-Talent Program, by the Key Research Program of Frontier Sciences, CAS, and by Project 11647601 supported by NSFC.


\begin{thebibliography}{99}
\bibitem{a}
  J.~M.~Maldacena,
  \textit{The Large N limit of superconformal field theories and supergravity},
  Int.\ J.\ Theor.\ Phys.\  {\bf 38} (1999) 1113,  [Adv.\ Theor.\ Math.\ Phys.\  {\bf 2} (1998) 231],   {\tt [arXiv:hep-th/9711200]}.

%\cite{Gubser:2002tv}
\bibitem{GKP} 
  S.~S.~Gubser, I.~R.~Klebanov and A.~M.~Polyakov,
  ``A Semiclassical limit of the gauge / string correspondence,''
  Nucl.\ Phys.\ B {\bf 636}, 99 (2002)
%  doi:10.1016/S0550-3213(02)00373-5
  [hep-th/0204051].
  %%CITATION = doi:10.1016/S0550-3213(02)00373-5;%%
  %955 citations counted in INSPIRE as of 10 Jun 2019
  
  %\cite{Kruczenski:2004wg}
\bibitem{spiky1} 
  M.~Kruczenski,
  ``Spiky strings and single trace operators in gauge theories,''
  JHEP {\bf 0508}, 014 (2005)
 % doi:10.1088/1126-6708/2005/08/014
  [hep-th/0410226].
  %%CITATION = doi:10.1088/1126-6708/2005/08/014;%%
  %160 citations counted in INSPIRE as of 10 Jun 2019
  
  %\cite{Kruczenski:2008bs}
\bibitem{Kruczenski:2008bs} 
  M.~Kruczenski and A.~A.~Tseytlin,
  ``Spiky strings, light-like Wilson loops and pp-wave anomaly,''
  Phys.\ Rev.\ D {\bf 77}, 126005 (2008)
  %doi:10.1103/PhysRevD.77.126005
  [arXiv:0802.2039 [hep-th]].
  %%CITATION = doi:10.1103/PhysRevD.77.126005;%%
  %42 citations counted in INSPIRE as of 10 Jun 2019
  
  %\cite{Ishizeki:2008tx}
\bibitem{Ishizeki:2008tx} 
  R.~Ishizeki, M.~Kruczenski, A.~Tirziu and A.~A.~Tseytlin,
  ``Spiky strings in AdS(3) x S1 and their AdS-pp-wave limits,''
  Phys.\ Rev.\ D {\bf 79}, 026006 (2009)
 % doi:10.1103/PhysRevD.79.026006
  [arXiv:0812.2431 [hep-th]].
  %%CITATION = doi:10.1103/PhysRevD.79.026006;%%
  %21 citations counted in INSPIRE as of 10 Jun 2019

%\cite{Jevicki:2008mm}
\bibitem{Jevicki:2008mm} 
  A.~Jevicki and K.~Jin,
  ``Solitons and AdS String Solutions,''
  Int.\ J.\ Mod.\ Phys.\ A {\bf 23}, 2289 (2008)
  %doi:10.1142/S0217751X0804113X
  [arXiv:0804.0412 [hep-th]].
  %%CITATION = doi:10.1142/S0217751X0804113X;%%
  %49 citations counted in INSPIRE as of 10 Jun 2019

  
  %\cite{Jevicki:2009uz}
\bibitem{JJ} 
  A.~Jevicki and K.~Jin,
  ``Moduli Dynamics of AdS(3) Strings,''
  JHEP {\bf 0906}, 064 (2009)
 % doi:10.1088/1126-6708/2009/06/064
  [arXiv:0903.3389 [hep-th]].
  %%CITATION = doi:10.1088/1126-6708/2009/06/064;%%
  %47 citations counted in INSPIRE as of 10 Jun 2019


  %\cite{Freyhult:2009bx}
\bibitem{Freyhult:2009bx} 
  L.~Freyhult, M.~Kruczenski and A.~Tirziu,
  ``Spiky strings in the SL(2) Bethe Ansatz,''
  JHEP {\bf 0907}, 038 (2009)
%  doi:10.1088/1126-6708/2009/07/038
  [arXiv:0905.3536 [hep-th]].
  %%CITATION = doi:10.1088/1126-6708/2009/07/038;%%
  %15 citations counted in INSPIRE as of 10 Jun 2019
  
  %\cite{Kruczenski:2010xs}
\bibitem{Kruczenski:2010xs} 
  M.~Kruczenski and A.~Tirziu,
  ``Spiky strings in Bethe Ansatz at strong coupling,''
  Phys.\ Rev.\ D {\bf 81}, 106004 (2010)
  %doi:10.1103/PhysRevD.81.106004
  [arXiv:1002.4843 [hep-th]].
  %%CITATION = doi:10.1103/PhysRevD.81.106004;%%
  %7 citations counted in INSPIRE as of 10 Jun 2019
  
  %\cite{Losi:2010hr}
\bibitem{Losi:2010hr} 
  M.~Losi,
  ``Spiky strings and the AdS/CFT correspondence,''
  %doi:10.17863/CAM.16106
  arXiv:1109.5401 [hep-th].
  %%CITATION = doi:10.17863/CAM.16106;%%
  %4 citations counted in INSPIRE as of 10 Jun 2019
  
  %\cite{Banerjee:2015nha}
\bibitem{Banerjee:2015nha} 
  A.~Banerjee, S.~Bhattacharya and K.~L.~Panigrahi,
  ``Spiky strings in $\varkappa$-deformed $AdS$,''
  JHEP {\bf 1506}, 057 (2015)
 % doi:10.1007/JHEP06(2015)057
  [arXiv:1503.07447 [hep-th]].
  %%CITATION = doi:10.1007/JHEP06(2015)057;%%
  %22 citations counted in INSPIRE as of 10 Jun 2019

\bibitem{b}
  J.~R.~David, G.~Mandal and S.~R.~Wadia,
  {\em Microscopic formulation of black holes in string theory},
  Phys.\ Rept.\  {\bf 369} (2002) 549,  {\tt [arXiv:hep-th/0203048]}.

%\cite{Sfondrini:2014via}
\bibitem{Sfondrini} 
  A.~Sfondrini,
  ``Towards integrability for ${\rm Ad}{{{\rm S}}_{{\bf 3}}}/{\rm CF}{{{\rm T}}_{{\bf 2}}}$,''
  J.\ Phys.\ A {\bf 48}, no. 2, 023001 (2015)
%  doi:10.1088/1751-8113/48/2/023001
  [arXiv:1406.2971 [hep-th]].
  %%CITATION = doi:10.1088/1751-8113/48/2/023001;%%
  %75 citations counted in INSPIRE as of 19 Jun 2018



\bibitem{VIII}
  J.~R.~David and B.~Sahoo,
  {\em Giant magnons in the D1-D5 system},
  JHEP {\bf 0807} (2008) 033,
  {\tt [arXiv:0804.3267 [hep-th]]}.
  
   J.~R.~David and B.~Sahoo,
  {\em S-matrix for magnons in the D1-D5 system},
  JHEP {\bf 1010} (2010) 112,
  {\tt [arXiv:1005.0501 [hep-th]]}.

\bibitem{SST} O.~Ohlsson Sax, B.~Stefa\'nski, Jr. and A.~Torrielli,
{\em On the massless modes of the AdS$_3$/CFT$_2$ integrable systems},
JHEP {\bf 1303} (2013) 109, {\tt [arXiv:1211.1952 [hep-th]]}.
%%CITATION = ARXIV:1211.1952;%%

\bibitem{AB} C.~Ahn and D.~Bombardelli,
{\em Exact $S$-matrices for AdS$_3$/CFT$_2$},
Int.\ J.\ Mod.\ Phys.\ A {\bf 28} (2013) 1350168, 
{\tt [arXiv:1211.4512 [hep-th]]}.
%%CITATION = ARXIV:1211.4512;%%

\bibitem{SW} P.~Sundin and L.~Wulff,
{\em Worldsheet scattering in AdS$_3$/CFT$_2$},
JHEP {\bf 1307} (2013) 007, {\tt [arXiv:1302.5349 [hep-th]]}.
%%CITATION = ARXIV:1302.5349;%%



\bibitem{Ads3S3T4A}
R.~Borsato, O.~Ohlsson Sax, A.~Sfondrini, B.~Stefański and A.~Torrielli,
{\em The all-loop integrable spin-chain for strings on AdS$_3 \times S^3 \times T^4$: the massive sector}, 
JHEP {\bf 1308} (2013) 043, 
{\tt [arXiv:1303.5995 [hep-th]]}.



\bibitem{Borsato:2013hoa} R.~Borsato, O.~Ohlsson Sax, A.~Sfondrini, B.~Stefa\'nski, Jr. and A.~Torrielli,
{\em Dressing phases of AdS3/CFT2}, 
Phys.\ Rev.\ D {\bf 88} (2013) 066004, {\tt [arXiv:1306.2512 [hep-th]]}.

\bibitem{Ads3S3T4B}R.~Borsato, O.~Ohlsson Sax, A.~Sfondrini and B.~Stefa\'nski, Jr.
{\em All-loop worldsheet S matrix for $AdS_3 \times S^3 \times T^4$}, 
Phys.\ Rev.\ Lett.\  {\bf 113} (2014) 13,  131601, 
{\tt [arXiv:1403.4543 [hep-th]]}.
%%CITATION = ARXIV:1403.4543;%%



\bibitem{BianchiHoare}  L.~Bianchi and B.~Hoare,
{\em $AdS_3 \times S^3 \times M^4$ string S-matrices from unitarity cuts},
JHEP {\bf 1408} (2014) 097, 
{\tt [arXiv:1405.7947 [hep-th]]}.
%%CITATION = ARXIV:1405.7947;%%

\bibitem{Spinconnection} R.~Borsato, O.~Ohlsson Sax, A.~Sfondrini and B.~Stefa\'nski, Jr.
{\em The complete $AdS_3 \times S^3 \times T^4$ worldsheet S-matrix}, 
JHEP {\bf 1410} (2014) 66, 
{\tt [arXiv:1406.0453 [hep-th]]}.
%%CITATION = ARXIV:1406.0453;%%


\bibitem{BothM4} A.~Pittelli, A.~Torrielli and M.~Wolf,
{\em Secret symmetries of type IIB superstring theory on $AdS_3 \times S^3 \times M^4$}, 
J.\ Phys.\ A {\bf 47} (2014) no.45,  455402, 
{\tt [arXiv:1406.2840 [hep-th]]}.


\bibitem{completeworldsheet} T.~Lloyd, O.~Ohlsson Sax, A.~Sfondrini and B.~Stefański, Jr.,
{\em The complete worldsheet S matrix of superstrings on $AdS_3 x S^3 x T^4$ with mixed three-form flux}, 
Nucl.\ Phys.\ B {\bf 891} (2015) 570, 
{\tt [arXiv:1410.0866 [hep-th]]}.


\bibitem{Ads3S3T4C}R.~Borsato, O.~Ohlsson Sax, A.~Sfondrini and B.~Stefański,
{\em On the spectrum of AdS$_3$ x S$^3$ x T$^4$ strings with Ramond–Ramond flux}, 
.\ Phys.\ A {\bf 49} (2016) no.41,  41LT03, {\tt [arXiv:1605.00518 [hep-th]]}.


\bibitem{BMN mismatch}
  P.~Sundin and L.~Wulff,
  {\em The complete one-loop BMN S-matrix in AdS$_{3} \times$S$^{3} \times$T$^{4}$},
  JHEP {\bf 1606} (2016) 062,
  {\tt[arXiv:1605.01632 [hep-th]]}.

\bibitem{Ads3S3T4D}R.~Borsato, O.~Ohlsson Sax, A.~Sfondrini, B.~Stefański, A.~Torrielli and O.~Ohlsson Sax,
{\em On the dressing factors, Bethe equations and Yangian symmetry of strings on AdS$_3 \times$ S$^3 \times$ T$^4$}, 
J.\ Phys.\ A {\bf 50} (2017) no.2,  024004, 
{\tt [arXiv:1607.00914 [hep-th]]}.

\bibitem{BPS}
  M.~Baggio, O.~Ohlsson Sax, A.~Sfondrini, B.~Stefański and A.~Torrielli,
  {\em Protected string spectrum in AdS$_{3}$/CFT$_{2}$ from worldsheet integrability},
  JHEP {\bf 1704} (2017) 091
  {\tt[arXiv:1701.03501 [hep-th]]}.



















\bibitem{SS} O.~Ohlsson~Sax and B.~Stefa\'nski,~Jr.,
{\em Integrability, spin-chains and the AdS$_3$/CFT$_2$ correspondence},
JHEP {\bf 1108} (2011) 029, {\tt [arXiv:1106.2558 [hep-th]]}.


\bibitem{Rughoonauth} N.~Rughoonauth, P.~Sundin and L.~Wulff,
{\em Near BMN dynamics of the $AdS_3 \times S^3 \times S^3 \times S^1$ superstring},
JHEP {\bf 1207} (2012) 159, {\tt [arXiv:1204.4742 [hep-th]]}.



\bibitem{SU11} R.~Borsato, O.~Ohlsson~Sax and A.~Sfondrini,
{\em A dynamic $su(1|1)^2$ S-matrix for AdS$_3$/CFT$_2$}, 
JHEP {\bf 1304} (2013) 113, {\tt [arXiv:1211.5119 [hep-th]]}.



\bibitem{BOS} R.~Borsato, O.~Ohlsson Sax and A.~Sfondrini,
{\em All-loop Bethe ansatz equations for AdS3/CFT2}, 
JHEP {\bf 1304} (2013) 116, 
{\tt [arXiv:1212.0505 [hep-th]]}.


 R.~Borsato, O.~Ohlsson Sax, A.~Sfondrini and B.~Stefański,
{\em The $\mathrm{AdS}_3\times \mathrm{S}^3\times \mathrm{S}^3\times\mathrm{S}^1$ worldsheet S matrix}, 
J.\ Phys.\ A {\bf 48} (2015) no.41,  415401, 
{\tt [arXiv:1506.00218 [hep-th]]}.

\bibitem{Abbott} M.~C.~Abbott,
{\em Comment on Strings in AdS3 x S3 x S3 x S1 at One Loop}, 
JHEP {\bf 1302} (2013) 102, 
{\tt [arXiv:1211.5587 [hep-th]]}.
%%CITATION = doi:10.1007/JHEP02(2013)102;%%

M.~C.~Abbott,
{\em The $AdS_{3} \times S^{3} \times S^{3} \times S^{1}$ Hern\'andez-L\'opez phases: a semiclassical derivation},
J.\  Phys.\ A {\bf 46} (2013) 445401, {\tt [arXiv:1306.5106 [hep-th]]}.
%%CITATION = ARXIV:1306.5106;%%

M.~C.~Abbott and I.~Aniceto,
{\em Macroscopic (and Microscopic) Massless Modes},
Nucl.\ Phys.\ B {\bf 894} (2015) 75, 
{\tt [arXiv:1412.6380 [hep-th]]}.
%%CITATION = ARXIV:1412.6380;%%


\bibitem{EGGL} L.~Eberhardt, M.~R.~Gaberdiel, R.~Gopakumar and W.~Li,
{\em BPS spectrum on AdS$_3\times $S$^3 \times $S$^3 \times $S$^1$}, 
JHEP {\bf 1703} (2017) 124, 
{\tt [arXiv:1701.03552 [hep-th]]}.



%
%\bibitem{LarsenMartinec} F.~Larsen and E.~J.~Martinec,
%{\em U(1) charges and moduli in the D1 - D5 system}, 
%JHEP {\bf 9906} (1999) 019, {\tt [arXiv:hep-th/9905064]}.









\bibitem{XIV}
  S.~Gukov, E.~Martinec, G.~W.~Moore and A.~Strominger,
  {\em The Search for a holographic dual to AdS(3) x S**3 x S**3 x S**1},
  Adv.\ Theor.\ Math.\ Phys.\  {\bf 9} (2005) 435,
  {\tt [arXiv:hep-th/0403090]}.



\bibitem{Tong} D.~Tong,
{\em The holographic dual of $AdS_{3} \times  S^{3} \times S^{3} \times S^{1}$}, 
JHEP {\bf 1404} (2014) 193, 
{\tt [arXiv:1402.5135 [hep-th]]}.

\bibitem{EberhardtGopakumar} L.~Eberhardt, M.~R.~Gaberdiel and W.~Li,
{\em A holographic dual for string theory on AdS$_{3}\times$S$^{3}\times$S$^{3}\times$S$^{1}$}, 
JHEP {\bf 1708} (2017) 111, 
{\tt [arXiv:1707.02705 [hep-th]]}.

M.~R.~Gaberdiel and R.~Gopakumar,
{\em Tensionless String Spectra on ${\rm AdS}_3$}, 
{\tt [arXiv:1803.04423 [hep-th]]}.














\bibitem{BSZ} A.~Babichenko, B.~Stefa\'nski, Jr. and K.~Zarembo,
{\em Integrability and the $AdS_3/CFT_2$ correspondence},
JHEP {\bf 1003} (2010) 058, {\tt [arXiv:0912.1723 [hep-th]]}.

\bibitem{wulff} P.~Sundin and L.~Wulff,
{\em Classical integrability and quantum aspects of the AdS(3) x S(3) x S(3) x S(1) superstring},
JHEP {\bf 1210} (2012) 109, {\tt [arXiv:1207.5531 [hep-th]]}.


\bibitem{review} N.~Beisert {\it et al.},
{\em Review of AdS/CFT Integrability: An Overview}, 
Lett.\ Math.\ Phys.\  {\bf 99} (2012) 3, 
{\tt [arXiv:1012.3982 [hep-th]]}.


%\bibitem{Integrability in the CFT}
%  O.~Ohlsson Sax, A.~Sfondrini and B.~Stefa\'nski,
%  {\em Integrability and the Conformal Field Theory of the Higgs branch},
%  JHEP {\bf 1506} (2015) 103,
%  {\tt[arXiv:1411.3676 [hep-th]]}.









\bibitem{CagnazzoZarembo} A.~Cagnazzo and K.~Zarembo,
{\em B-field in AdS(3)/CFT(2) Correspondence and Integrability}, 
JHEP {\bf 1211} (2012) 133, {\tt [arXiv:1209.4049 [hep-th]]}.

\bibitem{Moo}
  J.~M.~Maldacena and H.~Ooguri,
  {\em Strings in $AdS(3)$ and $SL(2,R)$ WZW model 1.: The Spectrum},
  J.\ Math.\ Phys.\  {\bf 42} (2001) 2929,
  {\tt [arXiv:hep-th/0001053]}.



\bibitem{HT} B.~Hoare and A.~A.~Tseytlin,
{\em On string theory on $AdS_3 \times S^3 \times T^4$ with mixed 3-form flux: Tree-level S-matrix},
Nucl.\ Phys.\ B {\bf 873} (2013) 682, {\tt [arXiv:1303.1037 [hep-th]]}.


B.~Hoare and A.~A.~Tseytlin,
{\em Massive S-matrix of $AdS_3 \times S^3 \times T^4$ superstring theory with mixed 3-form flux},
Nucl.\ Phys.\ B {\bf 873} (2013) 395, {\tt [arXiv:1304.4099 [hep-th]]}.


\bibitem{B.Hoare}
B.~Hoare, A.~Stepanchuk and A.~A.~Tseytlin,
{\em Giant magnon solution and dispersion relation in string theory in $AdS_3 \times S^3 \times T^4$ with mixed flux},
Nucl.\ Phys.\ B {\bf 879} (2014) 318, {\tt [arXiv:1311.1794 [hep-th]]}.


\bibitem{Lloyd} T.~Lloyd, O.~Ohlsson Sax, A.~Sfondrini and B.~Stefa\'nski, Jr.,
{\em The complete worldsheet S matrix of superstrings on $AdS_3 \times S^3 \times T^4$ with mixed three-form flux},
Nucl.\ Phys.\ B {\bf 891} (2015) 570, 
{\tt [arXiv:1410.0866 [hep-th]]}.


\bibitem{SW2} P.~Sundin and L.~Wulff,
{\em One- and two-loop checks for the $AdS_3 \times S^3 \times T^4$ superstring with mixed flux},
J.\ Phys.\ A {\bf 48} (2015) 10,  105402, 
{\tt [arXiv:1411.4662 [hep-th]]}.

\bibitem{M.Baggio}
  M.~Baggio and A.~Sfondrini,
  {\em Strings on NS-NS Backgrounds as Integrable Deformations},
  {[\tt arXiv:1804.01998 [hep-th]]}.

\bibitem{Modulispace}
  O.~Ohlsson Sax and B.~Stefanski,
  {\em Closed Strings and Moduli in $AdS_3/CFT_2$},
  {\tt[arXiv:1804.02023 [hep-th]]}.




  
  


%\bibitem{FirstQuantizing} S.~Frolov and A.~A.~Tseytlin,
%{\em Semiclassical quantization of rotating superstring in AdS(5) x S**5}, 
%JHEP {\bf 0206} (2002) 007, 
%{\tt [arXiv:hep-th/0204226]}.
%
% S.~Frolov and A.~A.~Tseytlin,
%{\em Multispin string solutions in AdS(5) x S**5}, 
%Nucl.\ Phys.\ B {\bf 668} (2003) 77, 
%{\tt [arXiv:hep-th/0304255]}.


%\bibitem{Quantizing} S.~Frolov and A.~A.~Tseytlin,
%{\em Quantizing three spin string solution in AdS(5) x S**5}, 
%JHEP {\bf 0307} (2003) 016, 
%{\tt [arXiv:hep-th/0306130]}.

%\bibitem{Ghostcontribution} I.~Y.~Park, A.~Tirziu and A.~A.~Tseytlin,
%{\em Spinning strings in $AdS_5 \times S^5$: One-loop correction to energy in $SL(2)$ sector},
%  JHEP {\bf 0503} (2005) 013,
%  %doi:10.1088/1126-6708/2005/03/013
%  {\tt[arXiv:hep-th/0501203]}.
%
%\bibitem{Quantizing2} N.~Beisert and A.~A.~Tseytlin,
%{\em On quantum corrections to spinning strings and Bethe equations}, 
%Phys.\ Lett.\ B {\bf 629} (2005) 102, 
%{\tt [arXiv:hep-th/0509084]}.
%
%
%
%
%
%\bibitem{Schafer} S.~Schäfer-Nameki, M.~Zamaklar and K.~Zarembo,
%{\em Quantum corrections to spinning strings in AdS(5) x S(5) and Bethe ansatz: A Comparative study}, 
%JHEP {\bf 0509} (2005) 051, 
%{\tt [arXiv:hep-th/0507189]}.
%
%
%
%\bibitem{Nameki} S.~Schäfer-Nameki,
%{\em Exact expressions for quantum corrections to spinning strings}, 
%Phys.\ Lett.\ B {\bf 639} (2006) 571, 
%{\tt [arXiv:hep-th/0602214]}.
%
%
%
%
%\bibitem{SchaferNameki} S.~Schäfer-Nameki, M.~Zamaklar and K.~Zarembo,
%{\em How Accurate is the Quantum String Bethe Ansatz?}, 
%JHEP {\bf 0612} (2006) 020, 
%{\tt [arXiv:hep-th/0610250]}.
%
%
%
%
%
%
%
%  
%\bibitem{dunne} M.~Beccaria, G.~V.~Dunne, V.~Forini, M.~Pawellek and A.~A.~Tseytlin,
%{\em Exact computation of one-loop correction to energy of spinning folded string in $AdS_5 x S^5$}, 
%J.\ Phys.\ A {\bf 43} (2010) 165402, 
%{\tt [arXiv:1001.4018 [hep-th]]}.
%
%
%M.~Beccaria, G.~V.~Dunne, G.~Macorini, A.~Tirziu and A.~A.~Tseytlin,
%{\em Exact computation of one-loop correction to energy of pulsating strings in $AdS_5 x S^5$}, 
%J.\ Phys.\ A {\bf 44} (2011) 015404, 
%{\tt [arXiv:1009.2318 [hep-th]]}.
%
%\bibitem{Forini} V.~Forini, V.~G.~M.~Puletti, M.~Pawellek and E.~Vescovi,
%{\em One-loop spectroscopy of semiclassically quantized strings: bosonic sector},
%J.\ Phys.\ A {\bf 48} (2015) no.8,  085401, {\tt[arXiv:1409.8674 [hep-th]]}.
%
%\bibitem{FinitegapReview} N.~Gromov, 
%{\em Integrability in AdS/CFT correspondence: Quasi-classical analysis}, 
%Journal\ Of\ Physics\ A-Mathematical\ And\ Theoretical {\bf 25} (2009) 254004.



\bibitem{FinitegapBabichenko} A.~Babichenko, A.~Dekel and O.~Ohlsson Sax,
{\em Finite-gap equations for strings on AdS$_{3}$ x S$^{3}$ x T$^{4}$ with mixed 3-form flux}, 
JHEP {\bf 1411} (2014) 122, {\tt [arXiv:1405.6087 [hep-th]]}.

\bibitem{masslessfinitegap} T.~Lloyd and B.~Stefański, Jr.,
{\em $AdS_3/CFT_2$, finite-gap equations and massless modes}, 
JHEP {\bf 1404} (2014) 179, {\tt [arXiv:1312.3268 [hep-th]]}.
















\bibitem{Rotating1} J.~R.~David and A.~Sadhukhan,
{\em Spinning strings and minimal surfaces in $AdS_3$ with mixed 3-form fluxes}, 
JHEP {\bf 1410} (2014) 49, 
{\tt [arXiv:1405.2687 [hep-th]]}.


\bibitem{both}
A.~Banerjee, K.~L.~Panigrahi and P.~M.~Pradhan, 
{\em Spiky strings on $AdS_3 \times S^3$ with NS-NS flux}, 
Phys.\ Rev.\ D {\bf 90} (2014) no.10,  106006,
{\tt [arXiv:1405.5497 [hep-th]]}.

\bibitem{C.Ahn}
  C.~Ahn and P.~Bozhilov,
 {\em String solutions in $AdS_3 \times S^3 \times T^4$ with NS-NS B-field},
  Phys.\ Rev.\ D {\bf 90} (2014) no.6,  066010,
  {\tt [arXiv:1404.7644 [hep-th]]}.

%\cite{Banerjee:2018goh}
\bibitem{Banerjee:2018goh} 
  A.~Banerjee, S.~Biswas and K.~L.~Panigrahi,
  ``On multi-spin classical strings with NS-NS flux,''
  JHEP {\bf 1808}, 053 (2018)
  %doi:10.1007/JHEP08(2018)053
  [arXiv:1806.10934 [hep-th]].
  %%CITATION = doi:10.1007/JHEP08(2018)053;%%
  %1 citations counted in INSPIRE as of 08 Jun 2019

\bibitem{Rotating2}
A.~Banerjee, K.~L.~Panigrahi and M.~Samal, 
{\em A note on oscillating strings in AdS$_{3}$ x S$^{3}$ with mixed three-form fluxes}, 
JHEP {\bf 1511} (2015) 133,
{\tt [arXiv:1508.03430 [hep-th]]}.

%\cite{Barik:2017opb}
\bibitem{Barik1} 
  S.~P.~Barik, K.~L.~Panigrahi and M.~Samal,
 ``Perturbations of Pulsating Strings,''
  arXiv:1708.05202 [hep-th].
  %%CITATION = ARXIV:1708.05202;%%
  %1 citations counted in INSPIRE as of 19 Jun 2018

\bibitem{multispike}
A.~Banerjee and A.~Sadhukhan, 
{\em Multi-spike strings in AdS$_{3}$ with mixed three-form fluxes}, 
JHEP {\bf 1605} (2016) 083,
{\tt [arXiv:1512.01816 [hep-th]]}.

\bibitem{stepanchuk}
A. Stepanchuk
{\em Aspects of integrability in string sigma-models},
https://spiral.imperial.ac.uk/bitstream/10044/1/28904/1/Stepanchuk-A-2015-PhD-Thesis.pdf


\bibitem{HN1} R.~Hern\'andez and J.~M.~Nieto, 
{\em Spinning strings in $AdS_3 \times S^3$ with NS-NS flux}, 
Nucl.\ Phys.\ B {\bf 888} (2014) 236, 
{\tt [arXiv:1407.7475 [hep-th]]}.


\bibitem{HN2}
R.~Hern\'andez and J.~M.~Nieto, 
{\em Elliptic solutions in the Neumann-Rosochatius system with mixed flux}, 
Phys.\ Rev.\ D {\bf 91} (2015) no.12,  126006, 
{\tt [arXiv:1502.05203 [hep-th]]}.

%\cite{Hernandez:2018gcd}
\bibitem{HN3} 
  R.~Hern�ndez, J.~M.~Nieto and R.~Ruiz,
  ``Pulsating strings with mixed three-form flux,''
  JHEP {\bf 1804}, 078 (2018)
  doi:10.1007/JHEP04(2018)078
  [arXiv:1803.03078 [hep-th]].
  %%CITATION = doi:10.1007/JHEP04(2018)078;%%
  %2 citations counted in INSPIRE as of 19 Jun 2018

%\cite{Nieto:2018jzi}
\bibitem{HN4} 
  J.~M.~Nieto and R.~Ruiz,
 ``One-loop quantization of rigid spinning strings in $AdS_3 \times S^3 \times T^4$ with mixed flux,''
  arXiv:1804.10477 [hep-th].
  %%CITATION = ARXIV:1804.10477;%%
  %1 citations counted in INSPIRE as of 19 Jun 2018



\bibitem{D1}
A.~Banerjee, S.~Biswas and R.~R.~Nayak, 
{\em D1 string dynamics in curved backgrounds with fluxes}, 
JHEP {\bf 1604} (2016) 172,
{\tt [arXiv:1601.06360 [hep-th]]}.

%%\cite{Ryang:2006yq}
%\bibitem{Ryang} 
%  S.~Ryang,
%  ``Three-spin giant magnons in $AdS(5) x S**5$,''
%  JHEP {\bf 0612}, 043 (2006)
%  doi:10.1088/1126-6708/2006/12/043
%  [hep-th/0610037].
%  %%CITATION = doi:10.1088/1126-6708/2006/12/043;%%
%  %46 citations counted in INSPIRE as of 26 Jun 2018

%\cite{Hernandez:2018lvh}
\bibitem{Hernandez:2018lvh} 
  R.~Hern�ndez, J.~M.~Nieto and R.~Ruiz,
  %``Minimal surfaces with mixed three-form flux,''
  Phys.\ Rev.\ D {\bf 99}, no. 8, 086003 (2019)
  %doi:10.1103/PhysRevD.99.086003
  [arXiv:1811.08294 [hep-th]].
  %%CITATION = doi:10.1103/PhysRevD.99.086003;%%
  
  %\cite{Dei:2018mfl}
\bibitem{purens2} 
  A.~Dei and A.~Sfondrini,
  ``Integrable spin chain for stringy Wess-Zumino-Witten models,''
  arXiv:1806.00422 [hep-th].
  JHEP {\bf 1807}, 109 (2018)
  %%CITATION = ARXIV:1806.00422;%%
  %1 citations counted in INSPIRE as of 20 Jun 2018

%%\cite{Minahan:2006bd}
%\bibitem{Minahan1} 
%  J.~A.~Minahan, A.~Tirziu and A.~A.~Tseytlin,
%  ``Infinite spin limit of semiclassical string states,''
%  JHEP {\bf 0608}, 049 (2006)
%  doi:10.1088/1126-6708/2006/08/049
%  [hep-th/0606145].
%  %%CITATION = doi:10.1088/1126-6708/2006/08/049;%%
%  %99 citations counted in INSPIRE as of 21 Jun 2018










%\bibitem{NR} G.~Arutyunov, S.~Frolov, J.~Russo and A.~A.~Tseytlin,
%{\em Spinning strings in $AdS_5 \times S^5$ and integrable systems},
%Nucl.\ Phys.\ B {\bf 671} (2003) 3, {\tt [arXiv:hep-th/0307191]}.
%
%
%G.~Arutyunov, J.~Russo and A.~A.~Tseytlin,
%{\em Spinning strings in $AdS_5 \times S^5$: New integrable system relations},
%Phys.\ Rev.\ D {\bf 69} (2004) 086009, {\tt [arXiv:hep-th/0311004]}.






%\bibitem{Characteristicfrequencies}  
%M.~Blau, M.~O'Loughlin, G.~Papadopoulos and A.~A.~Tseytlin,
%{\em Solvable models of strings in homogeneous plane wave backgrounds},
%  Nucl.\ Phys.\ B {\bf 673}, 57 (2003),
%  {\tt [arXiv:hep-th/0304198]}.
%  R.~R.~Metsaev,
%  {\em Type IIB Green-Schwarz superstring in plane wave Ramond-Ramond background},
%  Nucl.\ Phys.\ B {\bf 625} (2002) 70,
%  {\tt[arXiv:hep-th/0112044]}

\end{thebibliography}
\end{document}